\documentclass[preprint,prd,
superscriptaddress,nofootinbib,showpacs,tightenlines]{revtex4}
\usepackage{amssymb,latexsym}
\usepackage{amsmath,amsbsy}
\usepackage{epsfig,bm}
\usepackage{graphicx}
\usepackage{comment}
\DeclareMathOperator{\tr}{tr}
\DeclareMathOperator{\Tr}{Tr}

\DeclareFontFamily{OT1}{pzc}{}
\DeclareFontShape{OT1}{pzc}{m}{it}%
             {<-> s * [1.40] pzcmi7t}{}
\DeclareMathAlphabet{\mathscr}{OT1}{pzc}%
                                 {m}{it}

\begin{document}
\def\a{{\alpha}}
\def\b{{\beta}}
\def\d{{\delta}}
\def\D{{\Delta}}
\def\e{{\varepsilon}}
\def\g{{\gamma}}
\def\G{{\Gamma}}
\def\k{{\kappa}}
\def\l{{\lambda}}
\def\L{{\Lambda}}
\def\m{{\mu}}
\def\n{{\nu}}
\def\o{{\omega}}
\def\O{{\Omega}}
\def\S{{\Sigma}}
\def\s{{\sigma}}
\def\th{{\theta}}

\def\ol#1{{\overline{#1}}}

\def\Dslash{D\hskip-0.65em /}
\def\Dtslash{\tilde{D} \hskip-0.65em /}

\def\CPT{{$\chi$PT}}
\def\QCPT{{Q$\chi$PT}}
\def\PQCPT{{PQ$\chi$PT}}
\def\tr{\text{tr}}
\def\str{\text{str}}
\def\diag{\text{diag}}
\def\order{{\mathcal O}}

\def\meff{{m^2_{\text{eff}}}}

\def\Meff{{M_{\text{eff}}}}
\def\cF{{\mathcal F}}
\def\cS{{\mathcal S}}
\def\cC{{\mathcal C}}
\def\cE{{\mathcal E}}
\def\cB{{\mathcal B}}
\def\cT{{\mathcal T}}
\def\cQ{{\mathcal Q}}
\def\cL{{\mathcal L}}
\def\cO{{\mathcal O}}
\def\cA{{\mathcal A}}
\def\cV{{\mathcal V}}
\def\cR{{\mathcal R}}
\def\cH{{\mathcal H}}
\def\cW{{\mathcal W}}
\def\cM{{\mathcal M}}
\def\cD{{\mathcal D}}
\def\cN{{\mathcal N}}
\def\cP{{\mathcal P}}
\def\cK{{\mathcal K}}
\def\Qt{{\tilde{Q}}}
\def\Dt{{\tilde{D}}}
\def\psit{{\tilde{\psi}}}
\def\St{{\tilde{\Sigma}}}
\def\cBt{{\tilde{\mathcal{B}}}}
\def\cDt{{\tilde{\mathcal{D}}}}
\def\cTt{{\tilde{\mathcal{T}}}}
\def\cMt{{\tilde{\mathcal{M}}}}
\def\At{{\tilde{A}}}
\def\Qt{{\tilde{Q}}}
\def\cNt{{\tilde{\mathcal{N}}}}
\def\cOt{{\tilde{\mathcal{O}}}}
\def\cPt{{\tilde{\mathcal{P}}}}
\def\cI{{\mathcal{I}}}
\def\cJ{{\mathcal{J}}}

\def\eqref#1{{(\ref{#1})}}

\preprint{JLAB-THY-09-1111}
\preprint{UMD-40762-474}

\title{Extracting Nucleon Magnetic Moments and Electric Polarizabilities from Lattice QCD in Background Electric Fields}

\author{W.~Detmold}
\email[]{wdetmold@wm.edu}
\affiliation{%
Department of Physics, 
College of William and Mary, 
Williamsburg, Virginia 23187-8795,
USA
}
\affiliation{%
Thomas Jefferson National Accelerator Facility, 
Newport News, Virginia 23606,
USA
}

\author{B.~C.~Tiburzi}
\email[]{bctiburz@umd.edu}
\affiliation{%
Maryland Center for Fundamental Physics, 
Department of Physics, 
University of Maryland, 
College Park,  
Maryland 20742-4111, 
USA
}

\author{A.~Walker-Loud}
\email[]{walkloud@wm.edu}
\affiliation{%
Department of Physics, 
College of William and Mary, 
Williamsburg, Virginia 23187-8795,
USA
}

\date{\today}


\begin{abstract}
Nucleon properties are investigated in background electric fields. 
As the magnetic moments of baryons affect their relativistic propagation in constant electric fields, 
electric polarizabilities cannot be determined without knowledge of magnetic moments.
This is analogous to the experimental situation, 
for which determination of polarizabilities from the Compton amplitude
requires subtraction of Born terms. 
With the background field method, 
we devise combinations of nucleon correlation functions in constant electric fields that isolate magnetic moments
and electric polarizabilities.
Using an ensemble of anisotropic gauge configurations with dynamical clover fermions, 
we demonstrate how both observables can be determined from lattice QCD simulations in background electric fields.
We obtain results for the neutron and proton, 
however,
our study is currently limited to electrically neutral sea quarks. 
The value we extract for the nucleon isovector magnetic moment is comparable to those
obtained from measuring lattice three-point functions at similar pion masses. 
\end{abstract}
\maketitle

\section{Introduction}
\label{what?}

Understanding low-energy properties of hadrons directly from QCD remains a challenging endeavor.
In this low-energy regime, 
quark and gluon interactions must be treated non-perturbatively, 
ultimately resulting in their confinement into hadrons.
After three decades of dedicated work, 
lattice QCD has evolved into a tool to address quantitatively the non-perturbative dynamics underlying hadrons and their interactions, 
see~\cite{DeGrand:2006aa} for an overview.
Electromagnetic moments and multipole polarizabilities
are low-energy properties of hadrons with transparent physical meaning. 
These properties characterize the distribution of charge and magnetism within a hadron, 
and the response of the charge and magnetism distributions to external fields, respectively. 
Low-energy properties of hadrons can be described using an effective theory of QCD, 
based upon treating pseudoscalar mesons as the Goldstone modes arising from 
spontaneous chiral symmetry breaking.
A picture of hadrons emerges from chiral dynamics: 
that of a hadronic core surrounded by a pseudoscalar meson cloud. 
In part, the electromagnetic properties of hadrons encode 
the distribution of charged mesons, 
and the stiffness of the charged meson cloud.
Chiral dynamics consequently makes predictions for the form of electromagnetic observables. 
Confirming these predictions both experimentally and from the lattice will be a milestone in our understanding 
of non-perturbative QCD dynamics.

Computation of hadronic electromagnetic properties using lattice QCD can be accomplished in at least two different ways. 
The current insertion method, 
see e.g.~\cite{Martinelli:1987bh},
can be used to determine hadronic matrix elements of the electromagnetic current. 
This method is ideal for the computation of electromagnetic form factors, 
but is limited in the extraction of multipole moments due to the available lattice momentum, 
which, for periodic boundary conditions, is quantized in units of 
$2 \pi / L$, 
where 
$L$ 
is the size of the lattice.%
\footnote{
For isovector form factors, 
the restriction to quantized momentum transfer can be lifted by imposing isospin twisted boundary conditions on the quark fields~\cite{Tiburzi:2005hg,Tiburzi:2006px}. 
The only known method to handle the isoscalar contribution is to increase the lattice volume. 
}
For current lattice sizes, 
the determination of multipole moments relies on a long extrapolation to vanishing momentum transfer. 
For multipole polarizabilities, 
the temporal extent of current lattices also makes direct computation of the Compton scattering tensor infeasible. 
Were lattices long enough to allow the computation of matrix elements with two current insertions, 
the extraction of polarizabilities would still require a long extrapolation to zero momentum. 
Alternately, 
hadronic properties can be determined using the background field method~\cite{Fucito:1982ff,Martinelli:1982cb,Bernard:1982yu}. 
With this method, 
one determines lattice two-point functions in the presence of classical external fields. 
Observables are then determined from the variation of these hadronic correlators
with the strength of the external field.

In the current context, 
we focus our attention on properties of spin-half baryons in external electric fields.
Neutral hadrons in electric fields have been investigated with lattice QCD using the 
quenched approximation at pion masses greater than 
$500 \, \texttt{MeV}$~\cite{Fiebig:1988en,Christensen:2004ca}.
A recent calculation has explored neutron properties at lower pion masses with electrically neutral sea quarks%
~\cite{Alexandru:2009id}.
There has also been a fully dynamical calculation for the neutron in an electric field using a pion mass of 
$760 \, \texttt{MeV}$~\cite{Engelhardt:2007ub}.
In this work, 
we treat the baryon spin in a relativistic manner, 
and thereby demonstrate 
how to determine both nucleon magnetic moments and electric polarizabilities using lattice QCD in background electric fields.
In essence, 
our method allows one to perform the subtraction of the nucleon pole term arising from anomalous magnetic couplings.
The resulting polarizabilities consequently have the correct physical interpretation, 
and correspond to those extracted from experiment, 
as well as those derived from chiral perturbation theory, for example. 
We consider both the neutron and proton in this study. 
For the latter, 
we use the relativistic generalization of the method proposed in~%
\cite{Detmold:2006vu},
which relies on matching QCD correlators onto those derived from single-hadron effective actions.
We have also employed this method recently for pseudoscalar meson electric polarizabilities~\cite{Detmold:2009dx}.
A salient feature of these computations
is that they utilize a periodic lattice action with everywhere constant electric fields,
and thereby eliminate difficulties arising from Dirichlet boundary conditions used in previous studies.%
\footnote{
Periodic actions with everywhere constant magnetic fields 
have also been employed recently to study magnetic moments of hadrons~\cite{Aubin:2008qp}, 
and modification of the QCD vacuum~\cite{Buividovich:2009ih,Buividovich:2009wi}.
}
Our calculations of nucleon magnetic moments and electric polarizabilities 
include effects from dynamical quarks, 
however, 
they are restricted to electrically neutral sea quarks.

We organize our presentation in the following manner. 
First in Sect.~\ref{Correlators}, 
we analytically determine the form of baryon correlation functions in external electric fields. 
We specialize to the case of a uniform electric field, 
and derive results for both neutral and charged baryons. 
A key observation of this section is that baryon electric polarizabilities cannot be determined without
knowledge of their magnetic moments. 
This is analogous to the experimental situation, 
where Born terms must be subtracted to extract polarizabilities from Compton scattering. 
Appendix~\ref{a:A} is concerned with the physics underlying the Born subtraction. 
In Sect.~\ref{Details},
we provide the pertinent details of our lattice computations,
and implementation of the background field. 
In Sect.~\ref{Results}, 
we present our analysis of nucleon correlation functions 
calculated in background electric fields using lattice QCD. 
For both the neutron and proton, 
we demonstrate that the measured correlation functions
agree in form with the analytic expectations from the hadronic theory, 
and that magnetic moments and electric polarizabilities can be extracted from data. 
Appendix~\ref{a:B} is devoted to the analysis of unpolarized neutron correlators,    
from which consistent results are obtained.
(These results, however, do not permit the determination of the electric polarizability---only 
a combination of the polarizability and the square of the magnetic moment. 
The latter contribution arises from Born-level couplings.) 
A brief conclusion in Sect.~\ref{summy} ends our work.

\section{Spin-Half Correlation Functions}                                                  %
\label{Correlators}                                                                                             %

To extract properties of nucleons in background electric fields, 
we must first understand the expected behavior of their two-point correlation functions. 
In this section, 
we determine baryon two-point functions using the single-hadron effective action
that arises from QCD in the ultra-low energy limit. 
The functional forms deduced for these two-point functions can then be utilized
to fit baryon correlators computed with lattice QCD. 
From these fits, 
one can deduce hadronic parameters, 
such as the magnetic moments and electric polarizabilities.

To arrive at a uniform electric field of the form
$\vec{\cE} = \cE \hat{z}$,
we use the Euclidean space vector potential
\begin{equation}  \label{eq:A}
A_\mu = ( 0, 0, - \cE x_4, 0)
.\end{equation}
While there are other gauge equivalent choices, 
we find Eq.~\eqref{eq:A} particularly useful.
\footnote{
On a torus, 
many of the gauge equivalent choices in infinite volume are no longer equivalent,
but differ by their holonomy. 
External fields with non-vanishing holonomy lead to new interactions that are finite volume artifacts~\cite{Hu:2007eb,Tiburzi:2008pa,Detmold:2009fr}.   
As the current study is restricted to one lattice volume, 
we postpone the investigation of finite volume effects to future work.
}
The analytic continuation, 
$\cE \to - i E_\text{M}$, 
is needed to recover Minkowski space results.
As our interests lie only with quantities perturbative in the strength of the field, 
this analytic continuation can be performed trivially, see~\cite{Tiburzi:2008ma}.
A Euclidean formulation is natural from the point of view of lattice gauge theory simulations;
moreover, 
the Euclidean formulation removes instabilities due to non-perturbative effects,
i.e.~the Schwinger mechanism~\cite{Schwinger:1951nm}. 
With the vector potential specified, 
we can determine the baryon two-point functions. 
As neutral and charged baryons propagate differently in electric fields, 
we handle each separately.

\subsection{Neutral Spin-Half Baryons}                     %

We consider first the case of a neutral spin-half particle of mass
$M$
described by the field 
$\psi (x)$. 
The Euclidean space correlation function in the hadronic theory we denote by 
$G_{\a \b}(x_4, \cE)$, 
which is given by
\begin{equation}
G_{\a \b}(x_4, \cE)
=
\int d\bm{x}  
\,
\langle 0 | 
\psi_\a (x) \ol \psi_\b (0) 
| 0 \rangle_{\cE} 
,\end{equation}
where the subscript denotes that the correlation function is calculated in the background 
electric field. 
Integrating over all space projects the correlator onto vanishing three-momentum, 
which is a good quantum number. 
For lattice QCD with spatially periodic boundary conditions, 
a sum over lattice sites accomplishes the same thing.

The energy of the neutral particle, 
$E(\cE)$,
depends on the strength of the electric field. 
For weak fields, 
the energy has the expansion
\begin{equation}  \label{eq:weak}
E(\cE) = M + \frac{1}{2} 4 \pi  \a_E \cE^2 + \ldots
,\end{equation}
where 
$\a_E$ 
is the electric polarizability, 
and the ellipsis denotes higher-order terms in even powers of the field. 
The quadratic Stark shift is positive due to our Euclidean space treatment. 
The magnetic moment, 
$\mu$,
is also important, 
and this coupling is entirely anomalous. 
The single-hadron effective action for a neutral spin-half particle takes the form
\begin{equation} \label{eq:Bneutral}
S_E 
= 
\int d^4 x
\, 
\ol \psi (x)  
\left[
\rlap \slash \partial + E(\cE) 
- \frac{\mu(\cE)}{4 M}
\sigma_{\mu \nu} F_{\mu \nu}
\right]
\psi(x)
.\end{equation}
This action describes the dynamics of the neutron in the ultra-low energy limit of QCD. 
The electromagnetic field strength tensor is
$F_{\mu \nu} = \partial_\mu A_\nu - \partial_\nu A_\mu$. 
For a background electric field, 
$\sigma_{\mu \nu} F_{\mu \nu} = 2 \vec{K} \cdot \vec{\cE}$, 
where 
$\vec{K} = i \vec{\gamma} \gamma_4$
is the generator of boosts in the 
spin-half 
representation of the Lorentz group. 
The magnetic moment coupling has been written as 
$\cE$-dependent.  
In small fields,  
$\mu(\cE)$ 
has a perturbative expansion in even powers of the field, 
and satisfies the zero-field limit,
$\mu(0) = \mu$. 
Using the effective action in 
Eq.~\eqref{eq:Bneutral}
to determine the unpolarized two-point function, 
we arrive at
\begin{eqnarray} \label{eq:Naive}
\Tr [ 
\, G(x_4, \cE) 
\,
] 
&=&
Z(\cE)
\exp \left[ - x_4 \, E_{\text{eff}} (\cE) \right]
,\end{eqnarray}
where the effective energy, 
$E_{\text{eff}}(\cE)$,
depends on the magnetic moment, 
and is given by
\begin{eqnarray} 
E_{\text{eff}} (\cE) 
&=&
E(\cE) 
- 
\frac{\mu(\cE)^2 \cE^2}{8 M^3}
\notag \\
\label{eq:Eeff}
&=&
M + 
\frac{1}{2}  
\cE^2 
\left(
4 \pi  \a_E - \frac{\mu^2}{4 M^3} 
\right)
+ 
\ldots \,\,
.\end{eqnarray}
In the second line, 
we have retained terms in the effective energy
only up to second order in the electric field. 
There are two distinct contributions at that order. 
The first is the expected shift in energy due to the polarizability, 
while the second is the analogue of the nucleon pole term arising 
from Born-level couplings to the magnetic moment. 
Appendix~\ref{a:A} details the physics underlying this Born-like contribution. 
While the magnetic moment contribution arises from a relativistic effect, 
the resulting shift in baryon energy occurs at the same order as that due to the polarizability.
\footnote{
In the chiral limit, 
the contribution from the magnetic moment term is suppressed 
relative to the electric polarizability by a factor of 
$m_\pi / M$. 
This suppression, 
however, 
owes to the singular behavior of the polarizability in that limit,
namely
$\alpha_E \sim 1/ m_\pi$. 
Because our lattice pion mass is larger than physical, 
we will make no assumption about the dominance of the 
polarizability near the chiral limit. 
}
Appendix~\ref{a:B} is devoted to the extraction of 
$E_\text{eff} ( \cE)$, 
using unpolarized neutron correlation functions.  
The correlator in Eq.~\eqref{eq:Naive}, 
however, 
does not allow access to the electric polarizability without knowledge of the magnetic moment.
A background field analogue of the Born subtraction is needed.%

To extract both the magnetic moment and electric polarizability, 
we use the boost projection operators
\begin{equation}
\cP_\pm 
= \frac{1}{2} ( 1 \pm K_3 )
,\end{equation}
where 
$K_3$ 
is the boost operator is the 
$\hat{z}$-direction.
The boost-projected correlation functions are given by
\begin{eqnarray}
G_\pm(x_4, \cE) 
&\equiv&
\Tr [ 
\, 
\cP_\pm
G(x_4, \cE) 
\,
] 
\\
&=&
Z(\cE)
\left(
1 \pm \frac{ \cE \mu}{ 2 M^2}
\right)
\exp
\left[ 
- x_4 \, 
E_{\text{eff}} (\cE) 
\right] \label{eq:neutral}
.\end{eqnarray}   
With the additional electric field dependence present in the amplitude, 
one can separate the electric polarizability from the magnetic 
moment by simultaneously analyzing both boost-projected correlators. 
The method we employ to accomplish this will be detailed below.

\subsection{Charged Spin-Half Baryons}                               %

Consider now a spin-half baryon with charge $Q$.
The magnetic moment, $\mu$, is a sum of two terms, 
$\mu = Q + \tilde{\mu}$. 
The piece proportional to the charge is the Dirac magnetic moment, 
while that denoted by
$\tilde{\mu}$ 
is the anomalous magnetic moment. 
Including terms relevant for a uniform external field,
the relativistic single-particle action for a charged baryon
has the form
\begin{equation} \label{eq:BL}
S_E 
= 
\int d^4 x
\, \ol \psi (x)  
\left[
\Dslash + E(\cE) 
- \frac{\tilde{\mu}(\cE)}{4 M}
\sigma_{\mu \nu} F_{\mu \nu}
\right]
\psi(x)
,\end{equation}
where the electromagnetically gauge covariant derivative is
$D_\mu = \partial_\mu + i Q A_\mu$,
and 
higher-order terms in the field strength appear parametrically in the  
$\cE$-dependent 
couplings in Eq.~\eqref{eq:BL}. 
The parameter 
$E(\cE)$
is the charged particle's rest energy, 
which has the weak field expansion in Eq.~\eqref{eq:weak}.  
The anomalous magnetic moment coupling,  
$\tilde{\mu}(\cE)$,
has a weak field expansion in even powers of the field, 
and satisfies the zero-field relation,
$\tilde{\mu}(0) = \tilde{\mu}$. 
The action in Eq.~\eqref{eq:BL} describes the proton in the ultra-low energy regime of QCD.

As with the neutral baryons, 
it is beneficial to consider boost-projected correlation functions. 
For charged baryons,
these have the form
\begin{eqnarray} \label{eq:charged}
G_\pm(x_4, \cE) 
&=&
Z(\cE)
\left( 
1 \pm \frac{\tilde{\mu} \cE}{2 M^2}
\right)
D
\left( 
x_4, 
E_{\text{eff}}(\cE)^2 \mp Q \cE , 
\cE
\right)
,\end{eqnarray}
where the function 
$D(x_4, E^2, \cE)$
is the relativistic propagator function~\cite{Tiburzi:2008ma}
\begin{equation} \label{eq:scalar}
D(x_4, E^2, \cE)
=
\int_0^\infty
ds
\sqrt{\frac{Q \cE}{ 2 \pi \sinh ( Q \cE s)}}
\exp 
\left[ 
- \frac{1}{2} Q \cE x_4^2 \coth ( Q \cE s ) 
- \frac{1}{2} E^2 s
\right]
.\end{equation}
When the charge is set to zero, 
we recover the neutral baryon correlation functions in 
Eq.~\eqref{eq:neutral}. 
As one can see, 
to determine charged baryon electric polarizabilities, 
we must also deduce their anomalous magnetic moments.
This can be achieved utilizing both boost-projected correlation 
functions. 
These fits are more complicated than for neutral hadrons;
however,
the function in Eq.~\eqref{eq:scalar} also describes the propagation
of a charged scalar in an electric field,
and our previous study demonstrated that such fits can be carried out~\cite{Detmold:2009dx}.

\section{Lattice Details}                                                                                        %
\label{Details}                                                                                                        %

To demonstrate our method for extracting nucleon 
magnetic moments and electric polarizabilities from lattice two-point functions, 
we have employed an ensemble of anisotropic gauge
configurations with ($2+1$)-flavors of dynamical clover fermions~\cite{Edwards:2008ja,Lin:2008pr}.
Our ensemble consists of 
$200$ 
lattices of size 
$L^3 \times \beta = 20^3 \times 128$. 
After an initial 
$1000$ 
thermalization trajectories, 
the lattices were chosen 
from an ensemble of 
$7000$
spaced either by 
$20$ 
or
$40$
trajectories to minimize autocorrelations. 
The lattice spacing in the spatial directions is
$a_s = 0.123(3) \, \texttt{fm}$~\cite{Edwards:2008ja,Lin:2008pr},
with a non-perturbatively tuned 
anisotropy parameter of 
$\xi \equiv a_s / a_t = 3.5$,
where 
$a_t$ 
is the temporal lattice spacing. 
The finer temporal spacing is a crucial feature for this study, 
as it allows us to fit reliably more complicated functional forms for
the time-dependence of correlation functions.
For this ensemble, 
the renormalized strange quark mass 
is near the physical value, 
while the renormalized light quark mass 
leads to a pion mass of
$m_\pi \approx 390\, \texttt{MeV}$.

On each configuration, 
we compute at least 
$10$ 
propagators for each of the up, down, and strange quarks
with random spatial source locations.  
Multiple inversions were made efficient using the EigCG inverter implemented  
in the Chroma lattice field theory library%
~\cite{Stathopoulos:2007zi}. 
Interpolating fields at the source are generated from gauge-covariantly
Gaussian-smeared quark fields~\cite{Teper:1987wt,Albanese:1987ds} 
on a stout-smeared~\cite{Morningstar:2003gk}
gauge field in order to optimize the overlap onto the ground state in the absence of background fields. 
Interpolating fields at the sink are constructed from local quark fields.
Each propagator is located with source time at 
$(x_4)_{\text{src}} = 0$.  
Randomization of the source time location, 
while improving the statistical sampling, 
would complicate the determination of charged baryon correlation functions, 
as their two-point functions are no longer time-translationally invariant. 
The correlator given in Eq.~\eqref{eq:charged} 
is generally a function of the sink time-slice and not simply 
a function of the source-sink separation.
The full dependence on source time is given in~\cite{Tiburzi:2008ma}.

To implement the background field on the lattice, 
we modify the 
$SU(3)$ 
color gauge links,
$U_\mu(x)$, 
for each quark flavor by multiplying by the color-singlet Abelian links,
$U_\mu^{(\cE)} (x)$,
for the external field, 
namely
\begin{equation} \label{eq:ModifiedAbelianLink}
U_\mu (x) 
\longrightarrow
U_\mu (x) U_\mu^{(\cE)} (x)
U_\mu^{(\cE)_\perp}(x)
,\end{equation} 
where
$U_\mu^{(\cE)} (x) = \exp [ i q A_\mu(x) ]$,
with 
$q$
as the quark electric charge in units of 
$e>0$, 
and the vector potential is chosen to be:
$A_\mu(x) = ( 0, 0,  - \cE^{\text{latt}} x_4, 0)$,
as in Eq.~\eqref{eq:A}.
The additional transverse links are given by
\begin{equation}
U_\mu^{(\cE)_\perp}(x)
= 
\exp [ i q \cE^{\text{latt}} \beta x_3 \, \delta_{\mu 4} \, \delta_{x_4, \beta - 1} ]
.\end{equation}
These additional links together with the quantization condition for a torus~\cite{'tHooft:1979uj
}
\begin{equation} \label{eq:Quant}
\cE^{\text{latt}} 
= 
\frac{ 2 \pi n} { |q_d | \, \beta L}
,\end{equation}
ensure that the flux through every elementary plaquette is 
$q_d \, \cE^{\text{latt}}$ for down quarks, and 
$q_u \, \cE^{\text{latt}}$ for up quarks~\cite{Smit:1986fn,Rubinstein:1995hc}. 
In the quantization condition, 
$q_d = - 1/3$
is the electric charge of the down quark, 
and
$n$
must take on integer values.
The use of non-quantized fields have been investigated in~\cite{Detmold:2008xk}, 
where it was found that boundary gradients in the field strength can lead to energy shifts in the bulk of the lattice 
as large as the sought after shifts due to the polarizabilities. 
Notice that
$\cE^{\text{latt}}$ 
is given in lattice units, 
and the conversion to the physical electric field strength
$\cE$
is given by
\begin{equation}
\cE = (e \, a_t a_s)^{-1} \cE^{\text{latt}}
.\end{equation}
As the gauge field multiplication in Eq.~\eqref{eq:ModifiedAbelianLink}
is carried out on pre-existing gauge configurations, 
the sea quarks remain electrically neutral. 
This approximation is imposed because of computational restrictions.

Using Eq.~\eqref{eq:ModifiedAbelianLink}, 
we computed propagators for nine values of the field strength, 
$n$,
corresponding to the integer appearing in the quantization condition, Eq.~\eqref{eq:Quant}. 
We use 
$n=0$, 
which corresponds to a vanishing external field, 
as well as 
$n = \pm 1$, $\ldots$, $ \pm 4$. 
On these lattices, 
the expansion parameter governing the deformation of a hadron's pion cloud is given by~\cite{Tiburzi:2008ma}:
$\left( e \, \cE /m_\pi^2  \right)^2
= 
0.18 \, n^2$.
From the size of this parameter, 
we anticipate the need to include terms beyond quadratic order in the electric field expansion of hadron energies. 
In our analysis, 
we include terms up to quartic order. 
Larger lattices with their smaller allowed field strengths
 will be required for better control over systematic uncertainty relating to the electric field expansion of observables.

\section{Lattice Results}                                                                                              %
\label{Results}                                                                                                             %

Nucleon two-point functions were obtained for each source location on a given configuration. 
Results for multiple source locations on each configuration were then source averaged, 
yielding lattice correlation functions we denote by: 
$g_{\pm} (x_4, n)_i$. 
Here 
$i$ 
labels the configuration, 
and 
$\pm$ 
refers to the boost projection. 
This procedure was repeated for each value of the external field. 
We have performed multiple differing procedures to analyze the data, 
of which we detail only one method thoroughly in the text. 
Consistent results were obtained from the other procedures.


%
%
\begin{figure}
\epsfig{file=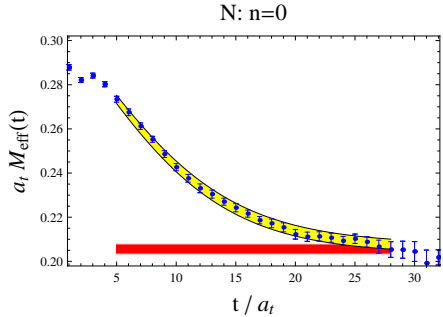,width=0.35\textwidth}
\caption{
Effective mass plot for the nucleon in vanishing electric field
($n = 0$). 
We also show the effective mass of a two-state fit to the lattice correlation function, 
where the error band reflects only the uncertainty in the ground state energy. 
The flat band shows the value of the extracted ground state energy with its uncertainty. 
}
\label{f:zerofield}
\end{figure}
%
%

To enforce invariance under parity transformations, 
for which 
$\cE \to - \cE$, 
we took the geometric mean of correlators calculated at 
$n$
and 
$-n$ 
on each configuration.
\footnote{
Consistent results were obtained by performing the analysis on the arithmetic mean of correlators. 
}
Specifically from the set of
$g_\pm (x_4, n)_i$, 
we form
\begin{eqnarray}
\mathfrak{g}_\pm(x_4, n)_i 
&=&
\sqrt{g_\pm (x_4, n)_i \, g_\mp ( x_4, -n)_i }
,\end{eqnarray}
for $n \geq 0$. 
This reduces the nine field values to five, 
corresponding to the integers
$n = 0, \ldots, 4$. 
This ensemble of correlation functions was then used to generate 
$200$
bootstrap ensembles. 
For the ensemble averaged correlation functions,
we use the same notation but without a configuration label,  
namely
$\mathfrak{g}_\pm(x_4, n)$.
Fits to the bootstrapped ensemble are performed as described below.

Fits to correlation functions in vanishing electric field are often guided by effective mass plots. 
Ordinarily one looks for a plateau in the effective mass, 
$M_{\text{eff}}(t)$, 
to ascertain when the excited state contributions have dropped out of the correlator. 
For the boost projected correlation functions, 
we define two different effective masses
\begin{equation}
M^\pm_{\text{eff}} (t) 
= 
- \log \frac{\mathfrak{g}_\pm(t+1, n) }{ \mathfrak{g}_\pm(t, n) 
}
.\end{equation}
The situation is quite simple for the case of vanishing electric field; 
thus we handle this case first.

When the external field vanishes, 
the two effective masses are identical, 
$M^+_{\text{eff}} (t) = M^-_{\text{eff}} (t) \equiv M_{\text{eff}} (t)$, 
and the standard analysis applies. 
In Fig.~\ref{f:zerofield}, 
we show the effective mass plot for the nucleon in zero electric field. 
Statistical noise dominates the correlator beyond the window of time depicted. 
As we are limited in statistics,
we perform a two-state fit to extract the mass of the ground state. 
The fit function, 
$\mathcal{G} (t, n =0)$, 
has the form
\begin{equation} \label{eq:fit000}
\mathcal{G} (t, 0) 
= 
Z(0) \exp ( - t M ) + Z'(0) \exp ( - t M')
,\end{equation}
where the parameters 
$Z(0)$, 
and 
$M$ 
arise from the ground state, 
while the primed parameters
account for excited state contamination. 
We use a correlated chi-squared analysis to fit the time dependence of the bootstrap
ensemble of correlation functions. 
As the amplitude parameters
$Z(0)$, 
and 
$Z'(0)$
enter the fit function linearly, 
we utilize variable projection
(see~\cite{Fleming:2004hs} for references)
to reduce the number of fit parameters from four down to two. 
The fit to the zero field nucleon correlation function has also been 
shown in the figure. 
The fit window has been determined by comparing single and double 
effective masses, see~\cite{Fleming:2004hs,Fleming:2009wb,Beane:2009ky} for details on the latter.
The ground state mass we extract from the two-state fit is consistent with the high statistics study%
~\cite{Beane:2009ky}.

\subsection{Neutron}                                    %

%
%
\begin{figure}
\epsfig{file=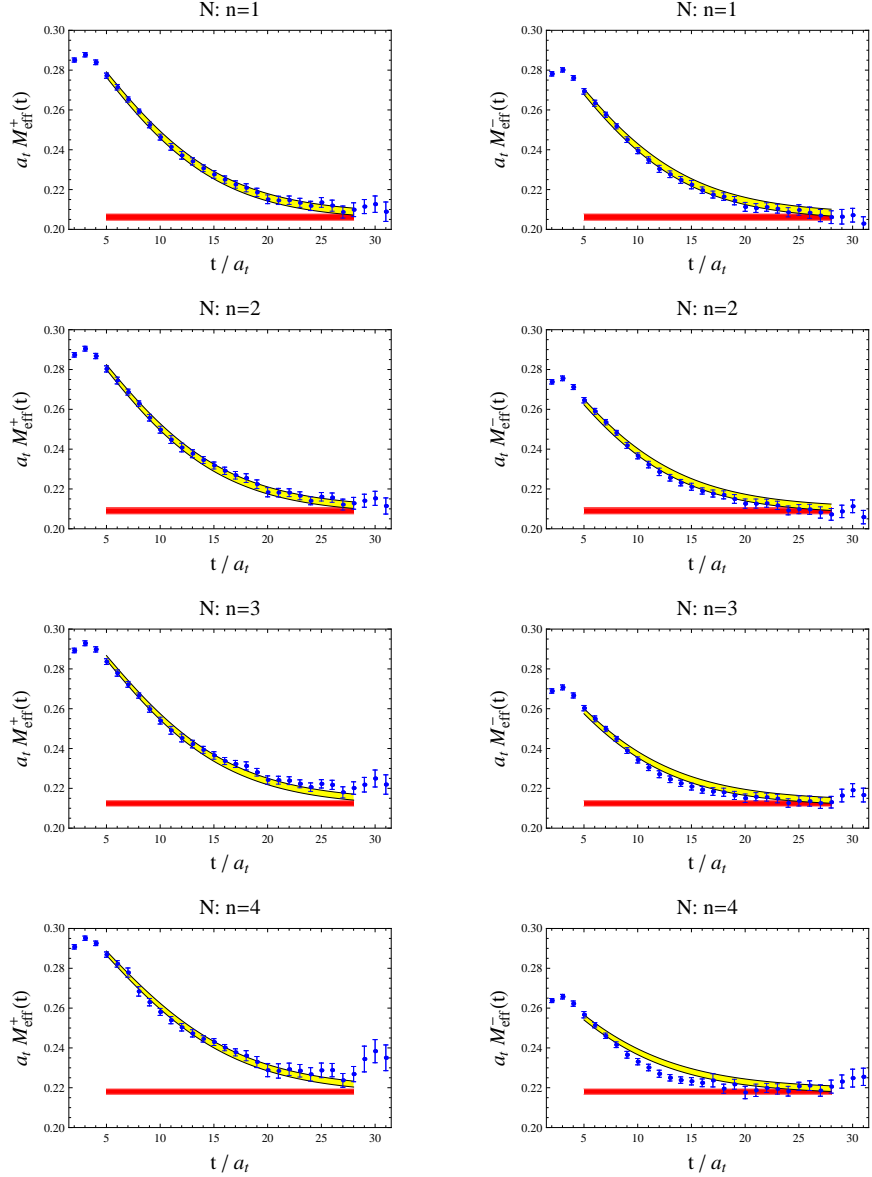,width=0.75\textwidth}
\caption{
Effective mass plots for the boost-projected neutron correlation functions.
For each value of the electric field strength, 
the curved band shows the result of the simultaneous fit to both boost-projected
correlation functions using Eq.~\eqref{eq:fit0}. 
The band accounts for the uncertainty in the extracted ground state energy, 
$E(n)$. 
The flat band shows the extracted value of
$E(n)$ 
with the uncertainty. 
}
\label{f:neutroneff}
\end{figure}
%
%

For non-vanishing electric fields, 
fit functions for the neutron and proton differ considerably.
For the neutron, 
the fit function is similar in form to the zero-field case, 
however, 
there are two distinct fit functions corresponding
to the boost projectors 
$\cP_\pm$, 
namely
\begin{eqnarray} \label{eq:fit0}
\mathcal{G}_\pm (t, n)
&=&
Z(n) 
\left[
1 \pm \mu^{\text{latt}} (n) \cE^{\text{latt}} / \xi
\right]
\exp 
\left[ 
- t \, E (n) \sqrt{1 - (\mu \, {}^{\text{latt}} (n) \cE^{\text{latt}}  / \xi )^2} 
\right]
\notag \\
&&
+ 
Z'(n)
\left[
1 \pm \mu'  \, {}^{\text{latt}} (n) \cE^{\text{latt}} / \xi
\right]
\exp 
\left[ 
- t \, E'  (n) \sqrt{1 - (\mu' \, {}^{\text{latt}} (n) \cE^{\text{latt}} / \xi)^2}  
\right]
.\end{eqnarray}
The unprimed parameters are those of the ground state, 
while the primed parameters account for excited state contributions. 
Notice both fit functions are identical to Eq.~\eqref{eq:fit000}
for vanishing electric field. 
For a fixed non-zero value of the electric field strength, 
there are six fit parameters, 
three for the ground state: 
$Z(n)$, 
$E(n)$, 
$\mu^{\text{latt}}(n)$, 
and similarly three for the excited state contribution.%
\footnote{
In principle, 
the amplitude 
$Z(n)$
may be different for the differing boost projections, 
$\mathcal{G}_+(t, n)$ 
and 
$\mathcal{G}_-(t,n)$.
Any such difference, 
however, 
is purely statistical in origin, 
and a suitable number of measurements should produce a common amplitude for the boost-projected correlators within uncertainties. 
} 
To perform the fits, 
variable projection is again utilized to remove the overall amplitudes,
$Z(n)$
and 
$Z'(n)$.
This reduces the number of fit parameters from six down to four. 
To determine the remaining four parameters, 
we perform simultaneous fits to both boost projected correlators
for each value of the external field.
\footnote{
A simultaneous fit is not required. 
Alternatively one can separately fit the two boost-projected correlation functions,
and combine these results to determine energies and magnetic moments. 
In pursuing this alternate procedure, 
we find that fits to the plus-projected correlation functions are always
better than fits to the minus-projected correlation functions. 
For small field strengths, 
the difference is insignificant; 
however, 
for the largest field strength,
the fit to the minus-projected correlation function is poor. 
We do not presently know the origin of this effect. 
Using a simultaneous fit to both boost-projected correlators mitigates (but does not remove) the problem with the largest field strength. 
}
In principle, 
such fits should take into account correlations between the 
boost projected correlators.
We find, 
however,
that the off-diagonal correlations between boost-projected correlators
are an order of magnitude smaller than the diagonal ones. 
Thus we treat the boost-projected correlators as uncorrelated, 
and fit them to the function in Eq.~\eqref{eq:fit0} taking into account correlations in time. 
In Fig.~\ref{f:neutroneff}, 
we show the effective mass plots for the boost projected neutron 
correlation functions. 
Along with these plots, 
we show the effective masses resulting from the simultaneous 
fit to both boost projected correlators. 
Details of the fits to correlation functions, and the extracted parameters are collected in Table~\ref{t:NeutralFit}. 
For the largest value of the field strength, 
$\cE$, 
the confidence of the fit is frankly poor, 
and so we are careful about using this data point.

%
\begin{table}
\caption{%
Summary of fit results for neutron two-point functions using the time window:
$5 \leq t / a_t \leq  28$.
All quoted values are averages over the bootstrap ensemble, 
and are given in dimensionless lattice units.
The conversion to physical units is detailed in the text. 
For the fits, 
$\chi^2 / d$
is the minimized value for chi-squared per degree of freedom, 
and 
$1-P$ 
is the integrated chi-squared from the minimum value to infinity.
The first half of the table summarizes the time-correlated fits to the energies and magnetic couplings in each field using 
Eq.~\eqref{eq:fit0}, 
while the second half summarizes the field-correlated fits
using Eqs.~\eqref{eq:energy} and \eqref{eq:magnetic}. 
The two differing fits to the latter are denoted by I and II, and are described in the text. 
The second uncertainty on polarizabilities and magnetic moments is an estimate
of the systematic due to the choice of fit window.
}
\begin{center}
\begin{tabular}{cccccc}
$N$ & $\quad n \quad $ & $a_t E(n)$ & $\quad \mu^{\text{latt}}(n) \quad$  & $\chi^2 / d$ & $\quad 1-P \quad$ 
\tabularnewline
\hline
\hline
& $0$ & $0.206(22)$ & -- & $0.55$ & $0.96$
\tabularnewline
& $1$ & $0.2074(16) $ &$-51(6)$ & $0.70$ & $0.93$ 
\tabularnewline
& $2$ & $0.2142(16) $ & $-52(3)$ & $0.91$ & $0.65$ 
\tabularnewline
& $3$ & $0.2240(15) $ & $-50(2)$ & $1.1$ & $0.24$
\tabularnewline
& $4$ & $0.2375(15)$ & $-47(1)$ & $1.5$ & $0.02$
\tabularnewline
\hline
\hline
\tabularnewline
\end{tabular}
\begin{tabular}{ccccc||ccc}
$N $ & $\quad  a_t M \quad $ &  $\quad \alpha_E^{\text{latt}} \quad  $  & $\chi^2 / d$ & $1-P $ &
$\quad \mu^{\text{latt}} \quad$  & $\chi^2 / d$ & $ 1-P $
\tabularnewline
\hline
I & $0.206(2)$ & $40(9)(2)$ & $0.3$ & $0.9$ &
$-52(2)(1)$ & $0.6$ & $0.7$ 
\tabularnewline
II & $0.205(2)$ & $42(19)(2)$ & $0.3$ & $0.9$ &
$-52(3)(1)$ & $0.7$ &  $0.6$
\tabularnewline
\hline
\hline
\end{tabular}
\end{center}
\label{t:NeutralFit}
\end{table}

The correlation function fits are carried out on each bootstrap ensemble. 
In particular, 
we arrive at an ensemble of ground state energies and ground state magnetic couplings for each 
magnitude of the electric field 
$\cE$, 
or equivalently the corresponding integer 
$n$.
These ensembles we generically denote by 
$\{ \cO_i (n) \}$, 
where $i$ indexes the bootstrap ensemble, 
$i = 1, \ldots N$, 
and 
$\cO$
represents either the ground state energy 
$E$, 
or magnetic coupling 
$\mu$. 
As the ensembles of configurations for different field strengths are generated
from the same underlying lattice configurations, 
correlations between the energies for different field strengths
will be significant and we account for these.
On the bootstrap ensemble of energies and magnetic couplings, 
we perform electric-field correlated fits  
to the function 
$O(n)$, 
where for the case of the ground state energy, 
$O(n) = E(n)$, 
with
\begin{equation}
\label{eq:energy}
E(n) 
= 
M 
+ 
\alpha_E^{\text{latt}} \, (\cE^{\text{latt}} )^2 
-  
\ol \alpha_{EEE}^{\text{latt}}  \, (\cE^{\text{latt}})^4
,\end{equation}
and for the case of the ground state magnetic coupling, 
$O(n) = \mu(n)$, 
with
\begin{equation}
\label{eq:magnetic}
\mu^{\text{latt}}(n)
=
\mu^{\text{latt}} 
+ 
\ol \mu^{\text{latt}}_{E} \, (\cE^{\text{latt}})^2
+
\ol \mu^{\text{latt}}_{EEE} \, (\cE^{\text{latt}})^4
.\end{equation}
With the ensemble average quantitites denoted by 
$\ol \cO(n) = \frac{1}{N} \sum_i \cO_i(n)$, 
we minimize the correlated chi-squared, 
namely 
\begin{equation}
\chi^2  \label{eq:Ecorr}
= 
\sum_{n, n'}
\Big[
\ol \cO(n) - O( n)
\Big]
C^{-1}_{n, n'}
\Big[
\ol \cO(n') - O( n')
\Big]
,\end{equation}
with the field-strength correlation matrix, 
$C_{n,n'}$, 
given by
\begin{equation}
C_{n, n'} 
= 
\frac{1}{N -1}
\sum_{i = 1}^{N}
\, \Big[ \ol \cO(n) - \cO_i (n) \Big] \,\Big[ \ol \cO(n') - \cO_i(n') \Big]
.\end{equation}
Because all fit parameters
enter the fit functions 
$O(n)$ 
linearly, 
the chi-squared minimization can be done analytically.  
Fits to the energy function are carried out on the bootstrap ensemble, 
resulting fit parameters are averaged, and the uncertainties from 
fitting and bootstrapping are added in quadrature. 
The same is done for the magnetic moment function, as defined in Eq.~\eqref{eq:magnetic}. 
We find that the best fits result from taking 
$\ol \mu^{\text{latt}}_{E} = 0$, 
and results quoted for the neutron use this constraint.
Furthermore we perform two different field-correlated fits as follows:
(I) a fit to all five field strengths using Eqs.~\eqref{eq:energy} and \eqref{eq:magnetic}, 
(II) the same fit function but excluding the largest field strength for which the quality of fit to the correlation functions is poor.
Finally, to estimate the systematics due to the choice of fit window, 
we performed uncorrelated fits to the electric field dependence 
of meson energies determined on adjacent fit windows. 
We chose the nine fit windows obtained by varying the start and end times
by one unit in either direction. 
On each time window, we determined the electric polarizability and magnetic moment.
The systematic uncertainty on 
these observables due to the fit window is estimated as the standard deviation of the extracted 
observables over the various adjacent windows.
Details of the correlated electric field fits and extracted parameters are tabulated in Table~\ref{t:NeutralFit}.

%
%
\begin{figure}
\epsfig{file=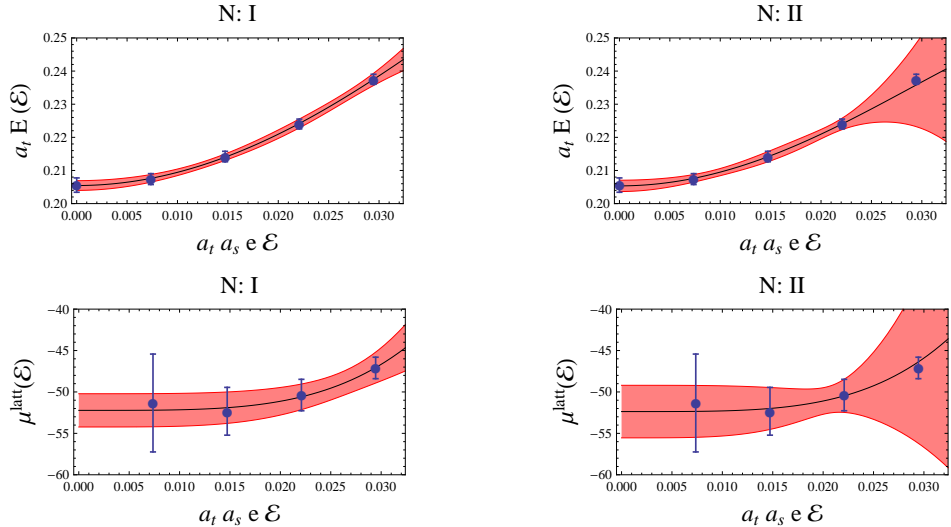,width=0.85 \textwidth}
\caption{
Electric field strength dependence of extracted neutron parameters.
The two different field-correlated fits [I (left panels) and II (right panels)] are described in the text. 
The bands in the plots reflect the total uncertainty.  
}
\label{f:neutronfield}
\end{figure}
%
%

From the extracted parameters, 
we can investigate the electric field dependence of the 
energies and magnetic moment couplings. 
This is done in Fig.~\ref{f:neutronfield}, 
where we plot the field strength dependence of these quantities. 
The plots, 
moreover, 
show the results of the two fits
(I and II) to the electric field dependence.
The values of the extracted parameters are consistent with na\"ive expectations
and it is useful to convert to physical units. 
Comparing the fit function in Eq.~\eqref{eq:fit0} to the correlator in physical units, Eq.~\eqref{eq:neutral}, 
we have 
\begin{equation} \label{eq:convertMU}
\mu = \frac{2 e \, ( a_t M)^3 }{  a_t M_N} \mu^{\text{latt}}
= 0.0313(7)  \times  \mu^{\text{latt}}
,\end{equation}
with the physical mangetic moment,
$\mu$,
given in units of nuclear magnetons, 
$\mu_N = \frac{e }{ 2 M_N}$, 
where 
$M_N$ 
is the physical mass of the nucleon, 
and the uncertainty arises from scale setting.
\footnote{  
Without a factor of 
$a_t M / a_t M_N$, 
the magnetic moment would be given in units of lattice nuclear magnetons, 
$\mu^{\text{latt}}_N = \frac{e}{2 M}$, 
with 
$M$
as the lattice value of the nucleon mass. 
With these units, 
there is no uncertainty from scale setting, 
however, 
they introduce additional pion mass dependence of the extracted moment. 
}
For the magnetic moment of the neutron in units of nuclear magnetons,
we thus find 
\begin{equation}
\mu_n^\text{conn} (m_\pi = 390 \, \texttt{MeV}) =  -1.63 (10)(4)(5) \, [\mu_N]
.\notag
\end{equation}  
We have appended a superscript to reflect that our computation includes 
only connected contributions. 
The three uncertainties quoted are from: 
(i) statistics and fitting, 
(ii) the systematic due to the fit window, 
and 
(iii) conversion to units of physical nuclear magnetons. 
For (i), 
we take the largerst value of the uncertainty from the two fits to the field-strength dependence
(which both gave the same value for $\mu^{\text{latt}}$). 
There are additional sources of systematic uncertainty that we have not unaccounted for,
namely the effects of finite lattice spacing and finite lattice volume.

To convert the lattice electric polarizability to physical units, 
we compare the fit function in lattice units, Eq.~\eqref{eq:energy}, 
to the energy in physical units, Eq.~\eqref{eq:weak}, and find
\begin{equation} \label{eq:convertALPHA}
\alpha_E =   \frac{e^2}{2 \pi}  a_t a_s^2 \, \alpha_E^{\text{latt}} = 0.0776 (58) \times \alpha_E^{\text{latt}} \times 10^{-4} \texttt{fm} {}^3
,\end{equation}
where the uncertainty arises from scale setting, 
and is specifically three times the uncertainty in the lattice spacing. 
For the neutron electric polarizability, 
we thus find
\begin{equation}
\alpha_E^n {}^\text{conn} (m_\pi = 390 \, \texttt{MeV}) =  3.3 (1.5)(2)(3) \times 10^{-4} \texttt{fm} {}^3
,\notag
\end{equation}
taking the central value and uncertainties from fit II. 
The notation and sources of uncertainty on the electric polarizability are as for the magnetic moment.

\subsection{Proton}                                      %

%
%
\begin{figure}
\epsfig{file=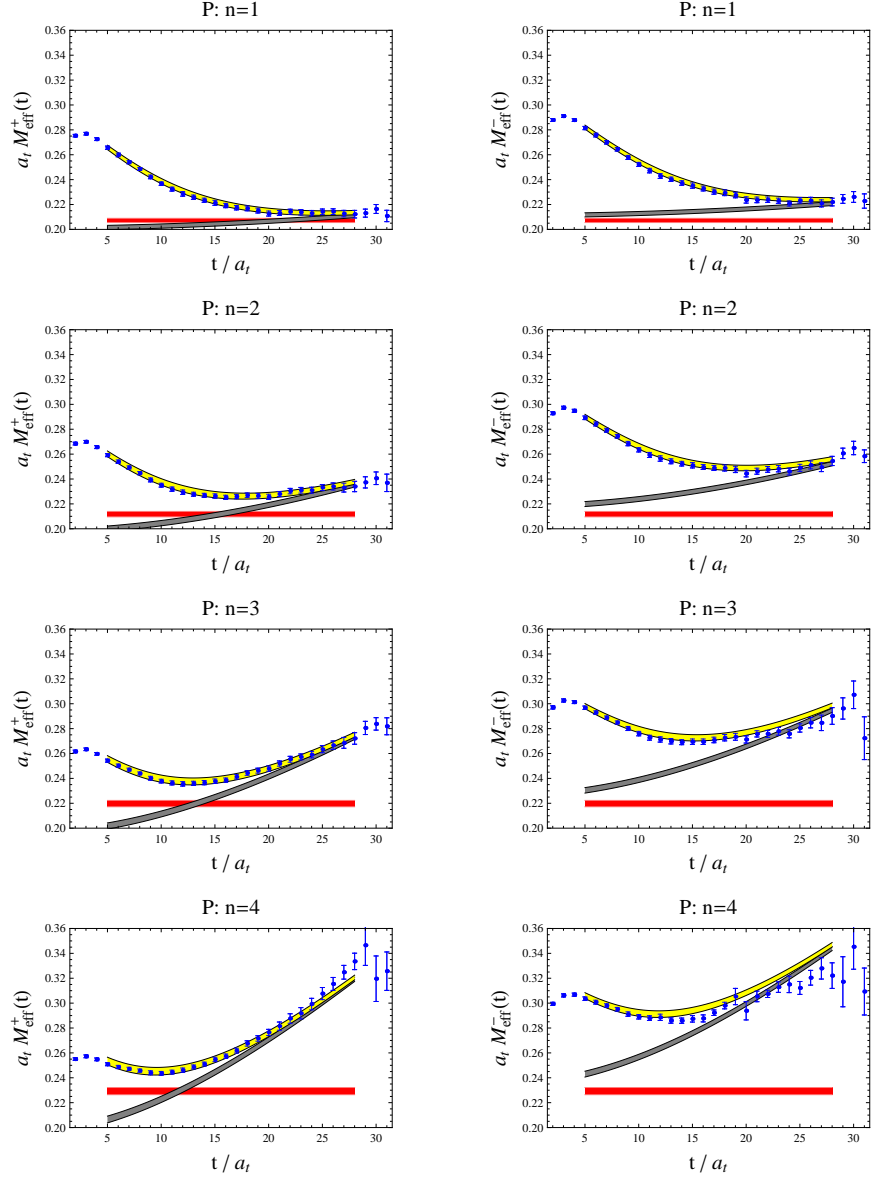,width=0.75\textwidth}
\caption{
Effective mass plots for the boost-projected proton correlation functions.
For each value of the electric field strength, 
the curved upper band shows the result of the simultaneous fit to both boost-projected
correlation functions using Eq.~\eqref{eq:fitQ}. 
The band accounts for the uncertainty in the extracted ground state rest energy, 
$E(\cE)$. 
The curved lower band shows the contribution to the fit from only the ground state, 
while the flat band shows just the extracted value of
$E(\cE)$ 
with its uncertainty. 
}
\label{f:protoneff}
\end{figure}
%
%

For the proton, 
we perform a similar analysis. 
Fits to proton correlation functions are carried out using a two-state fit function. 
This allows us to remove excited state contamination. 
We perform simultaneous time-correlated fits to both boost projected
proton correlation functions using the fit function
\begin{eqnarray} 
\mathcal{G}_\pm (t, n)
&=&
Z(n) 
\left[
1 \pm  \tilde{\mu}^{\text{latt}} (n) \, \frac{\cE^{\text{latt}} }{ \xi}
\right]
D
\left( 
t, 
E (n)^2 \Big[ 1 - \Big(  \tilde{\mu}^{\text{latt}} (n) \, \frac{\cE^{\text{latt}} }{ \xi} \Big)^2 \Big]  
\mp 
\frac{\cE^{\text{latt}}}{ \xi} , 
\frac{\cE^{\text{latt}} }{ \xi}
\right)
\notag \\
&& 
+ 
Z'(n)
\left[
1 \pm \tilde{\mu}'  \, {}^{\text{latt}} (n) \, \frac{\cE^{\text{latt}} }{ \xi}
\right]
D
\left( 
t, 
E' (n)^2 \Big[ 1 - \Big(  \tilde{\mu}' \, {}^{\text{latt}} (n)  \,  \frac{\cE^{\text{latt}} }{ \xi} \Big)^2 \Big]  
\mp 
\frac{\cE^{\text{latt}} }{ \xi}  , 
\frac{\cE^{\text{latt}} }{ \xi}   
\right)
, \notag \\ \label{eq:fitQ}
\end{eqnarray}
with 
$D(x,E^2,\cE)$ 
as the relativistic propagator function given in Eq.~\eqref{eq:scalar}.
As the overall amplitudes 
$Z(n)$ 
and
$Z'(n)$
enter the fit function linearly, 
we utilize variable projection to eliminate them from the simultaneous fits. 
For each value of the electric field 
$\cE$
(or equivalently the integer $n$),
there are then four parameters in the fit: 
the ground state rest energy, 
$E(n)$,
the ground state anomalous magnetic coupling,
$\tilde{\mu}(n)$, 
as well the rest energy and anomalous magnetic coupling 
for the excited state. 
Note that we force the proton charge to have the value 
$Q = 1$.
Because we have an improved current, 
we expect only 
$\cO(a^2)$
differences from the continuum value. 
In Fig.~\ref{f:protoneff}, 
we show the effective mass plots for the boost projected proton 
correlation functions. 
Along with these plots, 
we show the effective masses resulting from the simultaneous 
fit to both boost projected correlators using Eq.~\eqref{eq:fitQ}. 
Details of the fits to proton correlation functions, 
and the extracted parameters from the fits are collected in Table~\ref{t:ChargedFit}.

%
\begin{table}
\caption{%
Summary of fit results for proton two-point functions using the time window:
$6 \leq t / a_t \leq  28$.
Tabulated entries are as in Table~\ref{t:NeutralFit}, with the exception that we have 
denoted the magnetic couplings as anomalous using tildes.
}
\begin{center}
\begin{tabular}{cccccc}
$P$ & $\quad n \quad $ & $a_t E(n)$ & $\quad \tilde{\mu}^{\text{latt}}(n) \quad$  & $\chi^2 / d$ & $\quad 1-P \quad$ 
\tabularnewline
\hline
\hline
& $0$ & $0.2052(24)$ & -- & $0.61$ & $0.93$
\tabularnewline
& $1$ & $0.2072(17) $ &$53(6)$ & $0.65$ & $0.97$ 
\tabularnewline
& $2$ & $0.2118(22) $ & $48(3)$ & $0.81$ & $0.80$ 
\tabularnewline
& $3$ & $0.2198(26) $ & $46(2)$ & $0.86$ & $0.74$
\tabularnewline
& $4$ & $0.2293(29)$ & $41(1)$ & $1.5$ & $0.02$
\tabularnewline
\hline
\hline
\tabularnewline
\end{tabular}
\begin{tabular}{ccccc||ccc}
$P $ & $\quad  a_t M \quad $ &  $\quad \alpha_E^{\text{latt}} \quad  $  & $\chi^2 / d$ & $1-P $ &
$\quad \tilde{\mu}^{\text{latt}} \quad$  & $\chi^2 / d$ & $ 1-P $
\tabularnewline
\hline
I & $0.205(2)$ & $32(13)(1)$ & $0.14$ & $0.98$ &
$52(3)(1)$ & $1.3$ & $0.3$ 
\tabularnewline
II & $0.205(2)$ & $31(25)(4)$ & $0.16$ & $0.96$ &
$52(4)(1)$ & $1.7$ &  $0.2$
\tabularnewline
\hline
\hline
\end{tabular}
\end{center}
\label{t:ChargedFit}
\end{table}

We perform fits to proton correlations functions on the entire bootstrap ensemble. 
This enables us to form an ensemble of extracted parameters for each value of the field strength, 
In particular, 
we consider the ensemble of extracted ground state rest energies,
$ \{ E_i(n) \}$, 
and ground state anomalous magnetic moments,
$\{ \tilde{\mu}_i (n) \}$. 
Collectively we denote these ensembles by 
$\cO_i(n)$, 
with the ensemble average denoted by 
$\ol \cO(n)$. 
Electric field correlated fits are performed using the fit functions
in Eqs.~\eqref{eq:energy} and \eqref{eq:magnetic}.
For the latter it is the electric field dependence of the anomalous couplings that is being fit.  
Furthermore, 
we extract the anomalous magnetic moment using the constraint
$\ol \mu^{\text{latt}}_{EEE} = 0$, 
which results in better fits. 
For both observables, 
we perform the fit using all the data (fit I), 
and excluding results for the largest field strength (fit II). 
Results of the fits are collected in Table~\ref{t:ChargedFit}.
%
%
%
%
%
%
%
%
\begin{figure}
\epsfig{file=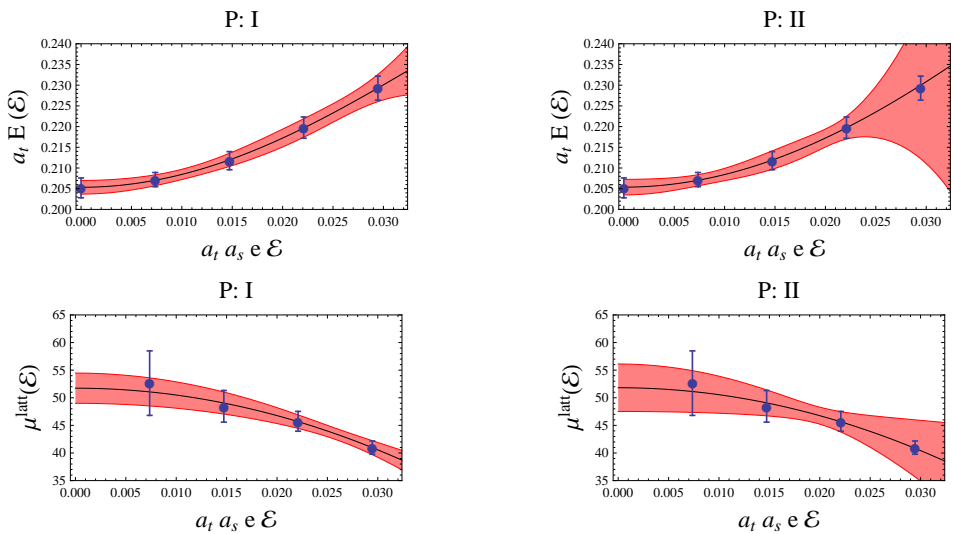,width=0.85\textwidth}
\caption{
Electric field strength dependence of extracted proton parameters.
Plots are as in Fig.~\ref{f:neutronfield}.
}
\label{f:protonfield}
\end{figure}
%
%
%
%
%
The electric field strength dependence of the extracted rest energies
and anomalous magnetic couplings are shown in Fig.~\ref{f:protonfield}. 
Also depicted are the field-correlated fits to these quantities.
The values of the extracted parameters are again consistent with na\"ive expectations. 
For the magnetic moment of the proton, 
converting to units of nuclear magnetons using Eq.~\eqref{eq:convertMU} yields
\begin{equation}
\mu_p^\text{conn} (m_\pi = 390 \, \texttt{MeV}) =  2.63 (13)(1)(4) \, [\mu_N]
.\notag
\end{equation}  
We have appended a superscript to reflect that our computation includes 
only connected contributions; we have also added in the Dirac contribution, 
assuming 
$Q = 1$. 
The three uncertainties quoted are as before:
(i) statistics and fitting, 
(ii) the systematic due to the fit window,
and 
(iii) conversion to physical units. 
For (i), 
we take the central value and uncertainty from fit II.  
For the proton electric polarizability,
converting the results of fit II to physical units using Eq.~\eqref{eq:convertALPHA} yields 
\begin{equation}
\alpha_E^p {}^\text{conn} (m_\pi = 390 \, \texttt{MeV}) =  2.4 (1.9)(3)(2) \times 10^{-4} \texttt{fm} {}^3
,\notag
\end{equation}
where the last uncertainty arises from scale setting.

Finally let us compare results for the neutron and proton. 
Within the uncertainty, 
the connected part of the isoscalar anomalous magnetic moment is consistent with zero. 
To compare with experiment, 
we require additional contributions from disconnected diagrams that we have not determined. 
The isovector combination of moments, 
however,   
does not have disconnected contributions due to strong isospin symmetry. 
For the nucleon isovector magnetic moment, 
we find
\begin{equation}
\mu_V (m_\pi = 390 \, \texttt{MeV}) 
= 4.3 (2)(1)(1) \,  [\mu_N]
.\notag
\end{equation} 
While this value is smaller than the physical moment, 
chiral corrections drive the magnetic moment downward at masses above the physical value~%
\cite{Bernard:1992qa,Bernard:1998gv}.
Studies at additional pion masses are necessary to 
extrapolate to the physical point. 
The value we obtain, 
moreover, 
is comparable to values extracted from the current insertion method at similar values of the pion mass, 
see~\cite{Hagler:2009ni}.

For the electric polarizabilities, 
our results show both isovector and isoscalar components, 
however, 
the latter is the dominant one. 
This is also seen experimentally and from chiral perturbation theory. 
The smaller isovector  component, 
\begin{equation}
\alpha_E^V (m_\pi = 390 \, \texttt{MeV}) = - 0.9(2.5)(3)(4) \times 10^{-4} \texttt{fm}^3
,\notag 
\end{equation}
receives smaller chiral corrections, 
and is less sensitive to the electric charges of the sea (but not independent). 
While values for the electric polarizabilities of the neutron and proton 
are smaller than experiment, 
chiral perturbation theory suggests large corrections as one nears the chiral limit~%
\cite{Bernard:1991rq,Bernard:1993bg,Hemmert:1996rw,Beane:2004ra}. 
Additionally including contributions from sea quark electric charges will drive
both polarizabilities upwards,
as can be seen from partially quenched chiral perturbation theory~\cite{Detmold:2006vu}. 
It will be interesting to carry out simulations at additional quark masses
and with electrically charged sea quarks to observe this behavior.

\section{Conclusion}                                                                                             %
\label{summy}                                                                                                         %

Above, we investigate the relativistic propagation of spin-half particles in classical electric fields. 
The presence of magnetic moments affects the behavior of two-point correlation functions,
and we use this observation to devise a method to determine magnetic moments and electric polarizabilities 
from lattice QCD simulated in background electric fields. 
Using anisotropic gauge configurations with dynamical clover fermions, 
we perform such computations. 
In Appendix~\ref{a:B}, 
we obtain results for the neutron using upolarized lattice correlation functions. 
Such results, 
however, 
do not allow one to determine the electric polarizability---only 
a combination of the electric polarizability and the square of the magnetic moment. 
The separation of these terms 
(which is analogous to accounting for Born-level contributions in the Compton scattering amplitude, 
see Appendix~\ref{a:A}) 
requires treatment of baryon spin, 
which is afforded by studying boost-projected correlators. 
Our analysis of boost-projected lattice correlation functions demonstrates that nucleon
magnetic moments and electric polarizabilities can be extracted from 
lattice calculations in background electric fields. 
This applies to both the neutron and proton.

There are a number of possible refinements of our computation that would reduce 
the systematic uncertainties. 
Currently,
our calculations are limited to electrically neutral sea quarks, 
and there is a need to remedy this situation. 
Furthermore, 
studies on larger volumes will not only reduce finite volume effects,
but allow the implementation of smaller values of the quantized 
external field strength. 
Calculations at various values of the quark mass will allow for chiral 
extrapolations to make contact with the physical QCD point. 
We intend to carry out this work in the future.
Finally, 
our approach can be used to study the magnetic moments 
and electric polarizabilities of the remaining members of the baryon octet.


\begin{acknowledgments}
These calculations were performed using code based on
the Chroma software suite~\cite{Edwards:2004sx}
on the computing clusters at Jefferson Laboratory.
Time on the clusters was awarded through the 
USQCD collaboration, and made possible by the SciDAC Initiative.
This work is supported in part by 
Jefferson Science Associates, LLC under 
U.S.~Dept.~of Energy contract No.~DE-AC05-06OR-23177 (W.D.).
The U.S. government retains a non-exclusive, paid-up
irrevocable, world-wide license to publish or reproduce
this mansuscript for U.S. government purposes. 
Additional support provided by the 
U.S.~Department of Energy, under
Grant Nos.~DE-SC000-1784 (W.D.),
~DE-FG02-93ER-40762 (B.C.T.), and
~DE-FG02-07ER-41527 (A.W.-L.).
\end{acknowledgments}


\appendix

\section{Motion-Induced Electric Dipole Moments}
\label{a:A}

Motion-induced electric dipole moments underly the Born-like terms encountered above. 
These Born couplings are magnetic in origin, 
and their subtraction is required to arrive at physical electric polarizabilities. 
Born subtractions are carried out when analyzing experimental data, 
calculating polarizabilities from chiral perturbation theory; 
and,
unique to this work,
determining electric polarizabilities from lattice QCD.
Here we remind the reader of the physics underlying the Born subtraction. 
For simplicity, 
we consider the neutron in an external electric field.
For the proton, 
there are additional Born couplings to the total charge. 
In momentum space, 
such terms can be subtracted; 
whereas in coordinate space,
these terms must be treated to all orders in the field strength.
This is accomplished by the relativistic proton propagator in Eq.~\eqref{eq:charged}.

One way to see that the Born term must be subtracted is to consider the electric dipole operator, 
$\vec{p}_E$. 
Consider the Minkowski space Hamiltonian derived from the Euclidean effective action for the neutron that appears in Eq.~\eqref{eq:Bneutral}.
The Minkowski space electric field we denote by 
$\vec{E}_M$.
Without subtracting the magnetic moment contribution, 
the electric dipole operator picks up an additional contribution
\begin{equation}
\vec{p}_E
= 
\frac{\partial H}{\partial \vec{E}_M}
= 
-  4 \pi \alpha_E \, \vec{E}_M
- \frac{\mu}{2 M} \vec{K}
,\end{equation}
that does not vanish when the electric field is turned off. 
This extra contribution is a motion-induced effect as can be seen from neutron matrix elements. 
For a neutron moving non-relativistically, 
\begin{equation}
\langle \vec{p}_E \rangle 
\equiv
\langle N(\vec{v}) | \, \vec{p}_E  \, | N (\vec{v}) \rangle
= 
- 4 \pi \alpha_E \, \vec{E}_M + \frac{\mu \langle \vec{\sigma} \rangle}{2 M}  \times \vec{v} + \ldots
,\end{equation}
we see that the additional term corresponds to a motion-induced dipole moment~\cite{Einstein:1906aa}. 
In the external field, 
the electric dipole moment contributes to the total energy in the form
\begin{equation}
\langle E \, \rangle
= 
\vec{E}_M
\cdot
\int_0^{E_M}
\langle \vec{p}_{E'} \rangle  \,
d E'_M
= 
- \frac{1}{2} 4 \pi \alpha_E \vec{E}_M^{\, 2}
+ 
\frac{\mu \langle \vec{\sigma} \rangle}{2 M}  
\cdot 
\left( 
\vec{v} \times \vec{E}_M 
\right)
.\end{equation}
The second term is readily identified as the interaction energy of the magnetic moment with  
magnetic field seen in the neutron's rest frame:
$ \langle \vec{m} \rangle \cdot \vec{B}$, 
with 
$\vec{m} = \frac{\mu \, \vec{\sigma} }{2 M}$, 
and 
$\vec{B} = \vec{v} \times \vec{E}_M$. 
This explains why one sees a motion-induced electric dipole moment 
in the frame in which neutron moves with velocity 
$\vec{v}$.

Without neutron motion, 
contributions from the motion-induced electric dipole moment na\"ively vanish. 
At second order, 
however, 
the motion-induced dipole can interact with itself via neutron propagation. 
We must employ a limiting procedure to handle the nucleon pole. 
In a quantum mechanical notation, 
the shift due to an intermediate-state neutron with energy 
$k_0 = M + \frac{1}{2} M \vec{v} \, {}^2 $ 
has the form
\begin{equation}
\Delta E
=
\Big
\langle N(\vec{0}) 
\Big|
\frac{\mu}{2 M} \vec{K} \cdot \vec{E}_M 
\Big| 
N (\vec{v}) 
\Big\rangle 
\frac{1}{M - k_0}
\Big\langle N(\vec{v}) \Big|
\frac{\mu}{2 M} \vec{K} \cdot \vec{E}_M 
\Big| N (\vec{0}) 
\Big\rangle 
.\end{equation}
The off-diagonal matrix elements in the non-relativistic limit
evaluate to half the value of the diagonal matrix elements in that limit. 
We can thus write the energy shift as
\begin{equation}
\Delta E 
= 
\frac{1}{4}
\vec{m} 
\cdot 
\left( 
\vec{v} \times \vec{E}_M 
\right)
\frac{1}{0 - \frac{1}{2} M \vec{v} \, {}^2} \;
\vec{m} 
\cdot 
\left( 
\vec{v} \times \vec{E}_M 
\right)
.\end{equation}
In the limit of zero velocity,
a non-vanishing contribution from the motion-induced electric dipole moment emerges. 
This contribution is the same as that derived in Eq.~\eqref{eq:Eeff}.

\section{Analysis of Unpolarized Neutron Correlation Functions} 
\label{a:B}

Here we present the analysis of unpolarized neutron correlation functions. 
On each configuration, 
for each value of the electric field strength, 
we form the unpolarized, 
source-averaged lattice correlation function,
$ \ol g(x_4, n)_i$.
Parity invariance is enforced by taking the geometric mean of 
correlators obtained for a given field value and its negative. 
Specifically we form
\begin{equation}
\overline{ \mathfrak{g}} (x_4, n)_i 
= 
\sqrt{ \ol g(x_4, n)_i  \, \ol g(x_4, - n)_i } 
,\end{equation}
for $n  \geq 0$. 
This ensemble of unpolarized correlators was then used to generate 
$200$ 
bootstrap ensembles for 
$n = 0, \ldots, 4$. 
The average unpolarized correlator is similarly denoted but without the subscript referring to configuration number, 
namely by 
$\overline{ \mathfrak{g}} (x_4, n)$.
The standard effective mass is then formed 
\begin{equation}
M_{\text{eff}} ( t) = - \log \frac{\overline{\mathfrak{g}} ( t+ 1, n) }{ \overline{\mathfrak{g}} ( t, n)}
,\end{equation}
and is used to guide spectroscopic analysis of the unpolarized correlators.

%
%
\begin{figure}
\epsfig{file=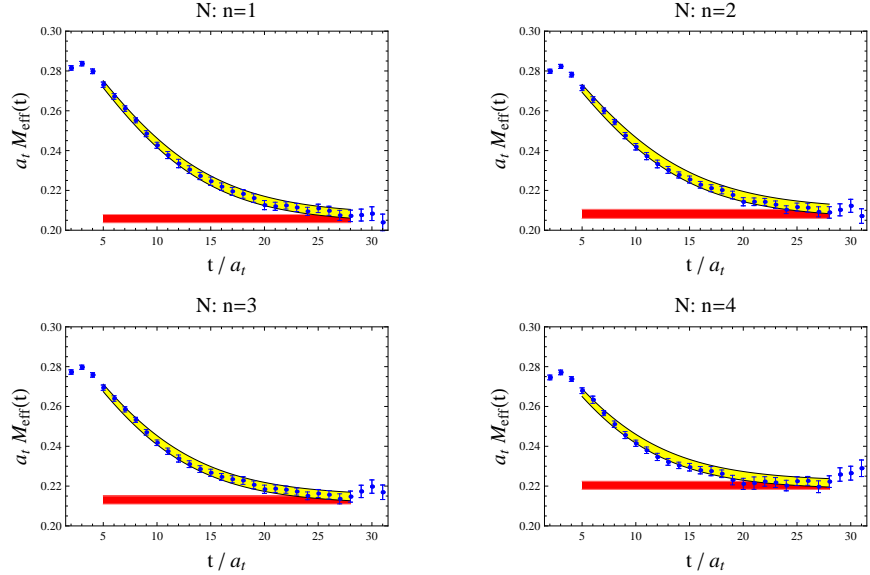,width=0.75\textwidth}
\caption{
Effective mass plots for unpolarized neutron correlation functions.
For each value of the electric field strength, 
the curved band shows the result of the fit to unpolarized correlation functions using Eq.~\eqref{eq:fit0A}. 
The band accounts for the uncertainty in the extracted ground-state effective energy, 
$E_{\text{eff}}(n)$. 
The flat band shows the extracted value of
$E_{\text{eff}}(n)$ 
with the uncertainty. 
}
\label{f:A_neutron}
\end{figure}
%
%

%
\begin{table}
\caption{%
Summary of fit results for unpolarized neutron two-point functions using the time window:
$5 \leq t / a_t \leq  28$.
All quoted values are averages over the bootstrap ensemble, 
and are given in dimensionless lattice units.
The quantity 
$\mathcal{A}_E^{\text{latt}}$ 
is the pseudo-polarizability defined in Eq.~\eqref{eq:PP}. 
For the fits, 
$\chi^2 / d$
is the minimized chi-squared per degree of freedom, 
and 
$1-P$ 
is the chi-squared integrated from the minimum value to infinity.
The first half of the table summarizes the time-correlated fits to the effective energies in each field using Eq.~\eqref{eq:fit0A}, 
while the second half summarizes the field-correlated fits using Eq.~\eqref{eq:energyA}. 
The two differing fits to the latter are denoted by I and II, 
and are described in the text. 
The second uncertainty on the pseudo-polarizability is an estimate
of the systematic due to the choice of fit window as explained in the text. 
}
\begin{center}
\begin{tabular}{ccccc}
$N$ & $\quad n \quad $ & $a_t E_{\text{eff}}(n)$  & $\quad \chi^2 / d \quad$ & $ 1-P $ 
\tabularnewline
\hline
\hline
& $0$ & $0.2041(23)$ & $0.50$ & $0.97$
\tabularnewline
& $1$ & $0.2058(21) $  & $0.67$ & $0.88$ 
\tabularnewline
& $2$ & $0.2082(23) $  & $0.97$ & $0.50$ 
\tabularnewline
& $3$ & $0.2130(21) $  & $0.70$ & $0.85$
\tabularnewline
& $4$ & $0.2204(21)$  & $0.70$ & $0.85$
\tabularnewline
\hline
\hline
\tabularnewline
\end{tabular}

\begin{tabular}{ccccc}
$\quad N \quad $ & $\quad  a_t M \quad $ &  $\qquad \mathcal{A}_E^{\text{latt}} \qquad  $  & $\chi^2 / d$ & $\quad 1-P \quad $ 
\tabularnewline
\hline
I & $0.205(2)$ & $17(11)(1)$ & $0.53$ & $0.75$ 
\tabularnewline
II & $0.204(2)$ & $20(25)(3)$ & $0.52$ & $0.72$ 
\tabularnewline
\hline
\hline
\end{tabular}
\end{center}
\label{t:A_neutron}
\end{table}

For a given value of the electric field, 
$\cE$, 
or equivalently the integer 
$n$, 
we extract the effective energy, 
$E_\text{eff} (n)$
given in Eq.~\eqref{eq:Eeff},
using a two-state fit function of the form
\begin{equation} \label{eq:fit0A}
\overline{\mathcal{G}} ( t, n) 
= 
Z( n) \exp \left[ - t E_\text{eff} ( n)  \right]
+ 
Z' (n) \exp \left[ - t E'_\text{eff} ( n) \right]
.\end{equation} 
Notice without the boost projection, 
we cannot disentangle the magnetic moment contribution. 
While there are four parameters to fit in Eq.~\eqref{eq:fit0A}, 
we utilize variable projection to eliminate the amplitudes
$Z(n)$ 
and 
$Z'(n)$, 
leaving just two parameters: 
the effective energy of the ground and excited states. 
Time-correlated fits are performed, 
with results shown in Fig.~\ref{f:A_neutron}.
Fit details and extracted parameters are collected in Table~\ref{t:A_neutron}.

%
%
\begin{figure}
\epsfig{file=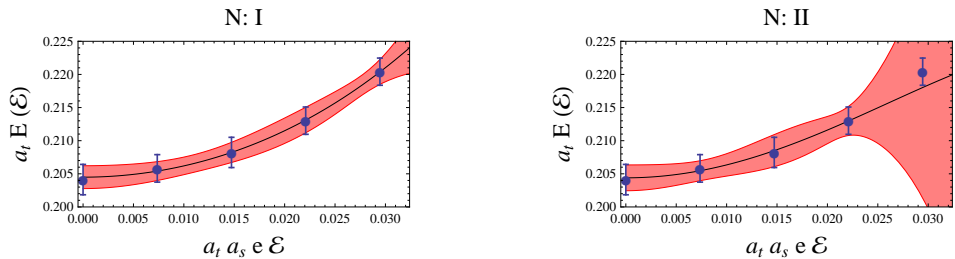,width=0.85 \textwidth}
\caption{
Electric field strength dependence of the neutron effective energy.
The two different field-correlated fits (I and II) are described in the text, 
and the bands show the total uncertainty. 
}
\label{f:A_neutronfield}
\end{figure}
%
%

Fits are carried out on the entire bootstrap ensemble 
enabling us to form an ensemble of extracted effective energies of the ground state, 
$\{ E_{\text{eff}, i} ( n) \}$.  
The average of this ensemble we denote
$\ol E_{\text{eff}} ( n) $.  
Using the fit function 
\begin{equation} \label{eq:energyA}
E_{\text{eff}} ( n ) 
= 
M + \mathcal{A}^{\text{latt}}_E \, ( \cE^{\text{latt}} )^2 + \mathcal{B}^{\text{latt}}_E \, ( \cE^{\text{latt}} )^4
,\end{equation}
electric field correlated fits are performed using 
Eq.~\eqref{eq:Ecorr}.
We perform two fits: a fit using all the data (fit I), 
and a fit that excludes results for the largest field strength (fit II). 
Results of the fits are collected in Table~\ref{t:A_neutron}.
The coefficient of the term quadratic in the field strength is not physically the electric polarizability 
because it includes Born-level contributions from the magnetic moment. 
For this reason, 
we call this coefficient the pseudo-polarizability.
In our choice of lattice units, 
the pseudo-polarizability is given by
\begin{equation} \label{eq:PP}
\mathcal{A}_E^{\text{latt}} 
=
\alpha_E^{\text{latt}}
- 
\frac{a_t M}{2 \xi^2} ( \mu^{\text{latt}} )^2
,\end{equation}
with 
$\xi$ 
as the anisotropy factor. 
The systematic due to the fit window is estimated by performing uncorrelated fits on the adjacent fit windows
obtained by varying the start and end times by one unit.

The extracted value for the pseudo-polarizability is roughly half the size of the electric polarizability, 
see Table~\ref{t:NeutralFit}.
Because of the sign of the magnetic moment contribution, 
we expect the pseudo-polarizability to be less than the electric polarizability. 
The values of the magnetic moment and electric polarizability extracted from boost-projected correlators
can be used to find a value for the pseudo-polarizability.
Using the results of Table~\ref{t:NeutralFit} in Eq.~\eqref{eq:PP}, 
we find
\begin{equation}
\text{I}: \; \mathcal{A}_E^{\text{latt}} = 17(9)(2), 
\quad  \text{and} \quad
\text{II}: \;  \mathcal{A}_E^{\text{latt}} = 19(19)(2)
,\end{equation}
for the two field-correlated fits.
These values are concordant with those found in Table~\ref{t:A_neutron} from analyzing the unpolarized neutron correlators.

Finally, 
we note that unpolarized proton correlation functions in principle allow one access 
to both the magnetic moment and electric polarizability. 
This can be seen from the explicit form for the proton two-point function derived in the
effective hadronic theory. 
The functional form, 
however, 
leads to fits that are challenging to perform. 
Valuable simplifications are afforded by boost projected correlators.

\bibliography{bibfile}

\begin{thebibliography}{43}
\expandafter\ifx\csname natexlab\endcsname\relax\def\natexlab#1{#1}\fi
\expandafter\ifx\csname bibnamefont\endcsname\relax
  \def\bibnamefont#1{#1}\fi
\expandafter\ifx\csname bibfnamefont\endcsname\relax
  \def\bibfnamefont#1{#1}\fi
\expandafter\ifx\csname citenamefont\endcsname\relax
  \def\citenamefont#1{#1}\fi
\expandafter\ifx\csname url\endcsname\relax
  \def\url#1{\texttt{#1}}\fi
\expandafter\ifx\csname urlprefix\endcsname\relax\def\urlprefix{URL }\fi
\providecommand{\bibinfo}[2]{#2}
\providecommand{\eprint}[2][]{\url{#2}}

\bibitem[{\citenamefont{DeGrand and DeTar}(2006)}]{DeGrand:2006aa}
\bibinfo{author}{\bibfnamefont{T.}~\bibnamefont{DeGrand}} \bibnamefont{and}
  \bibinfo{author}{\bibfnamefont{C.}~\bibnamefont{DeTar}},
  \emph{\bibinfo{title}{Lattice Methods for Quantum Chromodynamics}}
  (\bibinfo{publisher}{World Scientific}, \bibinfo{year}{2006}).

\bibitem[{\citenamefont{Martinelli and Sachrajda}(1988)}]{Martinelli:1987bh}
\bibinfo{author}{\bibfnamefont{G.}~\bibnamefont{Martinelli}} \bibnamefont{and}
  \bibinfo{author}{\bibfnamefont{C.~T.} \bibnamefont{Sachrajda}},
  \bibinfo{journal}{Nucl. Phys.} \textbf{\bibinfo{volume}{B306}},
  \bibinfo{pages}{865} (\bibinfo{year}{1988}).

\bibitem[{\citenamefont{Tiburzi}(2005)}]{Tiburzi:2005hg}
\bibinfo{author}{\bibfnamefont{B.~C.} \bibnamefont{Tiburzi}},
  \bibinfo{journal}{Phys. Lett.} \textbf{\bibinfo{volume}{B617}},
  \bibinfo{pages}{40} (\bibinfo{year}{2005}), \eprint{hep-lat/0504002}.

\bibitem[{\citenamefont{Tiburzi}(2006)}]{Tiburzi:2006px}
\bibinfo{author}{\bibfnamefont{B.~C.} \bibnamefont{Tiburzi}},
  \bibinfo{journal}{Phys.~Lett.} \textbf{\bibinfo{volume}{B641}},
  \bibinfo{pages}{342} (\bibinfo{year}{2006}), \eprint{hep-lat/0607019}.

\bibitem[{\citenamefont{Fucito et~al.}(1982)\citenamefont{Fucito, Parisi, and
  Petrarca}}]{Fucito:1982ff}
\bibinfo{author}{\bibfnamefont{F.}~\bibnamefont{Fucito}},
  \bibinfo{author}{\bibfnamefont{G.}~\bibnamefont{Parisi}}, \bibnamefont{and}
  \bibinfo{author}{\bibfnamefont{S.}~\bibnamefont{Petrarca}},
  \bibinfo{journal}{Phys. Lett.} \textbf{\bibinfo{volume}{B115}},
  \bibinfo{pages}{148} (\bibinfo{year}{1982}).

\bibitem[{\citenamefont{Martinelli et~al.}(1982)\citenamefont{Martinelli,
  Parisi, Petronzio, and Rapuano}}]{Martinelli:1982cb}
\bibinfo{author}{\bibfnamefont{G.}~\bibnamefont{Martinelli}},
  \bibinfo{author}{\bibfnamefont{G.}~\bibnamefont{Parisi}},
  \bibinfo{author}{\bibfnamefont{R.}~\bibnamefont{Petronzio}},
  \bibnamefont{and} \bibinfo{author}{\bibfnamefont{F.}~\bibnamefont{Rapuano}},
  \bibinfo{journal}{Phys. Lett.} \textbf{\bibinfo{volume}{B116}},
  \bibinfo{pages}{434} (\bibinfo{year}{1982}).

\bibitem[{\citenamefont{Bernard et~al.}(1982)\citenamefont{Bernard, Draper,
  Olynyk, and Rushton}}]{Bernard:1982yu}
\bibinfo{author}{\bibfnamefont{C.~W.} \bibnamefont{Bernard}},
  \bibinfo{author}{\bibfnamefont{T.}~\bibnamefont{Draper}},
  \bibinfo{author}{\bibfnamefont{K.}~\bibnamefont{Olynyk}}, \bibnamefont{and}
  \bibinfo{author}{\bibfnamefont{M.}~\bibnamefont{Rushton}},
  \bibinfo{journal}{Phys. Rev. Lett.} \textbf{\bibinfo{volume}{49}},
  \bibinfo{pages}{1076} (\bibinfo{year}{1982}).

\bibitem[{\citenamefont{Fiebig et~al.}(1989)\citenamefont{Fiebig, Wilcox, and
  Woloshyn}}]{Fiebig:1988en}
\bibinfo{author}{\bibfnamefont{H.~R.} \bibnamefont{Fiebig}},
  \bibinfo{author}{\bibfnamefont{W.}~\bibnamefont{Wilcox}}, \bibnamefont{and}
  \bibinfo{author}{\bibfnamefont{R.~M.} \bibnamefont{Woloshyn}},
  \bibinfo{journal}{Nucl. Phys.} \textbf{\bibinfo{volume}{B324}},
  \bibinfo{pages}{47} (\bibinfo{year}{1989}).

\bibitem[{\citenamefont{Christensen et~al.}(2005)\citenamefont{Christensen,
  Wilcox, Lee, and Zhou}}]{Christensen:2004ca}
\bibinfo{author}{\bibfnamefont{J.}~\bibnamefont{Christensen}},
  \bibinfo{author}{\bibfnamefont{W.}~\bibnamefont{Wilcox}},
  \bibinfo{author}{\bibfnamefont{F.~X.} \bibnamefont{Lee}}, \bibnamefont{and}
  \bibinfo{author}{\bibfnamefont{L.-M.} \bibnamefont{Zhou}},
  \bibinfo{journal}{Phys. Rev.} \textbf{\bibinfo{volume}{D72}},
  \bibinfo{pages}{034503} (\bibinfo{year}{2005}), \eprint{hep-lat/0408024}.

\bibitem[{\citenamefont{Alexandru and Lee}(2009)}]{Alexandru:2009id}
\bibinfo{author}{\bibfnamefont{A.}~\bibnamefont{Alexandru}} \bibnamefont{and}
  \bibinfo{author}{\bibfnamefont{F.~X.} \bibnamefont{Lee}}
  (\bibinfo{year}{2009}), \eprint{0911.2520}.

\bibitem[{\citenamefont{Engelhardt}(2007)}]{Engelhardt:2007ub}
\bibinfo{author}{\bibfnamefont{M.}~\bibnamefont{Engelhardt}}
  (\bibinfo{collaboration}{LHPC}), \bibinfo{journal}{Phys. Rev.}
  \textbf{\bibinfo{volume}{D76}}, \bibinfo{pages}{114502}
  (\bibinfo{year}{2007}), \eprint{0706.3919}.

\bibitem[{\citenamefont{Detmold et~al.}(2006)\citenamefont{Detmold, Tiburzi,
  and Walker-Loud}}]{Detmold:2006vu}
\bibinfo{author}{\bibfnamefont{W.}~\bibnamefont{Detmold}},
  \bibinfo{author}{\bibfnamefont{B.~C.} \bibnamefont{Tiburzi}},
  \bibnamefont{and}
  \bibinfo{author}{\bibfnamefont{A.}~\bibnamefont{Walker-Loud}},
  \bibinfo{journal}{Phys. Rev.} \textbf{\bibinfo{volume}{D73}},
  \bibinfo{pages}{114505} (\bibinfo{year}{2006}), \eprint{hep-lat/0603026}.

\bibitem[{\citenamefont{Detmold
  et~al.}(2009{\natexlab{a}})\citenamefont{Detmold, Tiburzi, and
  Walker-Loud}}]{Detmold:2009dx}
\bibinfo{author}{\bibfnamefont{W.}~\bibnamefont{Detmold}},
  \bibinfo{author}{\bibfnamefont{B.~C.} \bibnamefont{Tiburzi}},
  \bibnamefont{and}
  \bibinfo{author}{\bibfnamefont{A.}~\bibnamefont{Walker-Loud}},
  \bibinfo{journal}{Phys. Rev.} \textbf{\bibinfo{volume}{D79}},
  \bibinfo{pages}{094505} (\bibinfo{year}{2009}{\natexlab{a}}),
  \eprint{0904.1586}.

\bibitem[{\citenamefont{Aubin et~al.}(2009)\citenamefont{Aubin, Orginos,
  Pascalutsa, and Vanderhaeghen}}]{Aubin:2008qp}
\bibinfo{author}{\bibfnamefont{C.}~\bibnamefont{Aubin}},
  \bibinfo{author}{\bibfnamefont{K.}~\bibnamefont{Orginos}},
  \bibinfo{author}{\bibfnamefont{V.}~\bibnamefont{Pascalutsa}},
  \bibnamefont{and}
  \bibinfo{author}{\bibfnamefont{M.}~\bibnamefont{Vanderhaeghen}},
  \bibinfo{journal}{Phys. Rev.} \textbf{\bibinfo{volume}{D79}},
  \bibinfo{pages}{051502} (\bibinfo{year}{2009}), \eprint{0811.2440}.

\bibitem[{\citenamefont{Buividovich et~al.}(2010)\citenamefont{Buividovich,
  Chernodub, Luschevskaya, and Polikarpov}}]{Buividovich:2009ih}
\bibinfo{author}{\bibfnamefont{P.~V.} \bibnamefont{Buividovich}},
  \bibinfo{author}{\bibfnamefont{M.~N.} \bibnamefont{Chernodub}},
  \bibinfo{author}{\bibfnamefont{E.~V.} \bibnamefont{Luschevskaya}},
  \bibnamefont{and} \bibinfo{author}{\bibfnamefont{M.~I.}
  \bibnamefont{Polikarpov}}, \bibinfo{journal}{Nucl. Phys.}
  \textbf{\bibinfo{volume}{B826}}, \bibinfo{pages}{313} (\bibinfo{year}{2010}),
  \eprint{0906.0488}.

\bibitem[{\citenamefont{Buividovich et~al.}(2009)\citenamefont{Buividovich,
  Chernodub, Luschevskaya, and Polikarpov}}]{Buividovich:2009wi}
\bibinfo{author}{\bibfnamefont{P.~V.} \bibnamefont{Buividovich}},
  \bibinfo{author}{\bibfnamefont{M.~N.} \bibnamefont{Chernodub}},
  \bibinfo{author}{\bibfnamefont{E.~V.} \bibnamefont{Luschevskaya}},
  \bibnamefont{and} \bibinfo{author}{\bibfnamefont{M.~I.}
  \bibnamefont{Polikarpov}}, \bibinfo{journal}{Phys. Rev.}
  \textbf{\bibinfo{volume}{D80}}, \bibinfo{pages}{054503}
  (\bibinfo{year}{2009}), \eprint{0907.0494}.

\bibitem[{\citenamefont{Hu et~al.}(2007)\citenamefont{Hu, Jiang, and
  Tiburzi}}]{Hu:2007eb}
\bibinfo{author}{\bibfnamefont{J.}~\bibnamefont{Hu}},
  \bibinfo{author}{\bibfnamefont{F.-J.} \bibnamefont{Jiang}}, \bibnamefont{and}
  \bibinfo{author}{\bibfnamefont{B.~C.} \bibnamefont{Tiburzi}},
  \bibinfo{journal}{Phys. Lett.} \textbf{\bibinfo{volume}{B653}},
  \bibinfo{pages}{350} (\bibinfo{year}{2007}), \eprint{arXiv:0706.3408
  [hep-lat]}.

\bibitem[{\citenamefont{Tiburzi}(2009)}]{Tiburzi:2008pa}
\bibinfo{author}{\bibfnamefont{B.~C.} \bibnamefont{Tiburzi}},
  \bibinfo{journal}{Phys. Lett.} \textbf{\bibinfo{volume}{B674}},
  \bibinfo{pages}{336} (\bibinfo{year}{2009}), \eprint{0809.1886}.

\bibitem[{\citenamefont{Detmold
  et~al.}(2009{\natexlab{b}})\citenamefont{Detmold, Tiburzi, and
  Walker-Loud}}]{Detmold:2009fr}
\bibinfo{author}{\bibfnamefont{W.}~\bibnamefont{Detmold}},
  \bibinfo{author}{\bibfnamefont{B.~C.} \bibnamefont{Tiburzi}},
  \bibnamefont{and}
  \bibinfo{author}{\bibfnamefont{A.}~\bibnamefont{Walker-Loud}}
  (\bibinfo{year}{2009}{\natexlab{b}}), \eprint{0908.3626}.

\bibitem[{\citenamefont{Tiburzi}(2008)}]{Tiburzi:2008ma}
\bibinfo{author}{\bibfnamefont{B.~C.} \bibnamefont{Tiburzi}},
  \bibinfo{journal}{Nucl. Phys.} \textbf{\bibinfo{volume}{A814}},
  \bibinfo{pages}{74} (\bibinfo{year}{2008}), \eprint{0808.3965}.

\bibitem[{\citenamefont{Schwinger}(1951)}]{Schwinger:1951nm}
\bibinfo{author}{\bibfnamefont{J.~S.} \bibnamefont{Schwinger}},
  \bibinfo{journal}{Phys. Rev.} \textbf{\bibinfo{volume}{82}},
  \bibinfo{pages}{664} (\bibinfo{year}{1951}).

\bibitem[{\citenamefont{Edwards et~al.}(2008)\citenamefont{Edwards, Joo, and
  Lin}}]{Edwards:2008ja}
\bibinfo{author}{\bibfnamefont{R.~G.} \bibnamefont{Edwards}},
  \bibinfo{author}{\bibfnamefont{B.}~\bibnamefont{Joo}}, \bibnamefont{and}
  \bibinfo{author}{\bibfnamefont{H.-W.} \bibnamefont{Lin}},
  \bibinfo{journal}{Phys. Rev.} \textbf{\bibinfo{volume}{D78}},
  \bibinfo{pages}{054501} (\bibinfo{year}{2008}), \eprint{0803.3960}.

\bibitem[{\citenamefont{Lin et~al.}(2009)}]{Lin:2008pr}
\bibinfo{author}{\bibfnamefont{H.-W.} \bibnamefont{Lin}} \bibnamefont{et~al.}
  (\bibinfo{collaboration}{Hadron Spectrum}), \bibinfo{journal}{Phys. Rev.}
  \textbf{\bibinfo{volume}{D79}}, \bibinfo{pages}{034502}
  (\bibinfo{year}{2009}), \eprint{0810.3588}.

\bibitem[{\citenamefont{Stathopoulos and Orginos}(2007)}]{Stathopoulos:2007zi}
\bibinfo{author}{\bibfnamefont{A.}~\bibnamefont{Stathopoulos}}
  \bibnamefont{and} \bibinfo{author}{\bibfnamefont{K.}~\bibnamefont{Orginos}}
  (\bibinfo{year}{2007}), \eprint{0707.0131}.

\bibitem[{\citenamefont{Teper}(1987)}]{Teper:1987wt}
\bibinfo{author}{\bibfnamefont{M.}~\bibnamefont{Teper}},
  \bibinfo{journal}{Phys. Lett.} \textbf{\bibinfo{volume}{B183}},
  \bibinfo{pages}{345} (\bibinfo{year}{1987}).

\bibitem[{\citenamefont{Albanese et~al.}(1987)}]{Albanese:1987ds}
\bibinfo{author}{\bibfnamefont{M.}~\bibnamefont{Albanese}} \bibnamefont{et~al.}
  (\bibinfo{collaboration}{APE}), \bibinfo{journal}{Phys. Lett.}
  \textbf{\bibinfo{volume}{B192}}, \bibinfo{pages}{163} (\bibinfo{year}{1987}).

\bibitem[{\citenamefont{Morningstar and Peardon}(2004)}]{Morningstar:2003gk}
\bibinfo{author}{\bibfnamefont{C.}~\bibnamefont{Morningstar}} \bibnamefont{and}
  \bibinfo{author}{\bibfnamefont{M.~J.} \bibnamefont{Peardon}},
  \bibinfo{journal}{Phys. Rev.} \textbf{\bibinfo{volume}{D69}},
  \bibinfo{pages}{054501} (\bibinfo{year}{2004}), \eprint{hep-lat/0311018}.

\bibitem[{\citenamefont{'t~Hooft}(1979)}]{'tHooft:1979uj}
\bibinfo{author}{\bibfnamefont{G.}~\bibnamefont{'t~Hooft}},
  \bibinfo{journal}{Nucl. Phys.} \textbf{\bibinfo{volume}{B153}},
  \bibinfo{pages}{141} (\bibinfo{year}{1979}).

\bibitem[{\citenamefont{Smit and Vink}(1987)}]{Smit:1986fn}
\bibinfo{author}{\bibfnamefont{J.}~\bibnamefont{Smit}} \bibnamefont{and}
  \bibinfo{author}{\bibfnamefont{J.~C.} \bibnamefont{Vink}},
  \bibinfo{journal}{Nucl. Phys.} \textbf{\bibinfo{volume}{B286}},
  \bibinfo{pages}{485} (\bibinfo{year}{1987}).

\bibitem[{\citenamefont{Rubinstein et~al.}(1995)\citenamefont{Rubinstein,
  Solomon, and Wittlich}}]{Rubinstein:1995hc}
\bibinfo{author}{\bibfnamefont{H.~R.} \bibnamefont{Rubinstein}},
  \bibinfo{author}{\bibfnamefont{S.}~\bibnamefont{Solomon}}, \bibnamefont{and}
  \bibinfo{author}{\bibfnamefont{T.}~\bibnamefont{Wittlich}},
  \bibinfo{journal}{Nucl. Phys.} \textbf{\bibinfo{volume}{B457}},
  \bibinfo{pages}{577} (\bibinfo{year}{1995}), \eprint{hep-lat/9501001}.

\bibitem[{\citenamefont{Detmold et~al.}(2008)\citenamefont{Detmold, Tiburzi,
  and Walker-Loud}}]{Detmold:2008xk}
\bibinfo{author}{\bibfnamefont{W.}~\bibnamefont{Detmold}},
  \bibinfo{author}{\bibfnamefont{B.~C.} \bibnamefont{Tiburzi}},
  \bibnamefont{and}
  \bibinfo{author}{\bibfnamefont{A.}~\bibnamefont{Walker-Loud}}
  (\bibinfo{year}{2008}), \eprint{0809.0721}.

\bibitem[{\citenamefont{Fleming}(2004)}]{Fleming:2004hs}
\bibinfo{author}{\bibfnamefont{G.~T.} \bibnamefont{Fleming}}
  (\bibinfo{year}{2004}), \eprint{hep-lat/0403023}.

\bibitem[{\citenamefont{Fleming et~al.}(2009)\citenamefont{Fleming, Cohen, Lin,
  and Pereyra}}]{Fleming:2009wb}
\bibinfo{author}{\bibfnamefont{G.~T.} \bibnamefont{Fleming}},
  \bibinfo{author}{\bibfnamefont{S.~D.} \bibnamefont{Cohen}},
  \bibinfo{author}{\bibfnamefont{H.-W.} \bibnamefont{Lin}}, \bibnamefont{and}
  \bibinfo{author}{\bibfnamefont{V.}~\bibnamefont{Pereyra}}
  (\bibinfo{year}{2009}), \eprint{0903.2314}.

\bibitem[{\citenamefont{Beane et~al.}(2009)}]{Beane:2009ky}
\bibinfo{author}{\bibfnamefont{S.~R.} \bibnamefont{Beane}}
  \bibnamefont{et~al.}, \bibinfo{journal}{Phys. Rev.}
  \textbf{\bibinfo{volume}{D79}}, \bibinfo{pages}{114502}
  (\bibinfo{year}{2009}), \eprint{0903.2990}.

\bibitem[{\citenamefont{Bernard et~al.}(1992)\citenamefont{Bernard, Kaiser,
  Kambor, and Meissner}}]{Bernard:1992qa}
\bibinfo{author}{\bibfnamefont{V.}~\bibnamefont{Bernard}},
  \bibinfo{author}{\bibfnamefont{N.}~\bibnamefont{Kaiser}},
  \bibinfo{author}{\bibfnamefont{J.}~\bibnamefont{Kambor}}, \bibnamefont{and}
  \bibinfo{author}{\bibfnamefont{U.~G.} \bibnamefont{Meissner}},
  \bibinfo{journal}{Nucl. Phys.} \textbf{\bibinfo{volume}{B388}},
  \bibinfo{pages}{315} (\bibinfo{year}{1992}).

\bibitem[{\citenamefont{Bernard et~al.}(1998)\citenamefont{Bernard, Fearing,
  Hemmert, and Meissner}}]{Bernard:1998gv}
\bibinfo{author}{\bibfnamefont{V.}~\bibnamefont{Bernard}},
  \bibinfo{author}{\bibfnamefont{H.~W.} \bibnamefont{Fearing}},
  \bibinfo{author}{\bibfnamefont{T.~R.} \bibnamefont{Hemmert}},
  \bibnamefont{and} \bibinfo{author}{\bibfnamefont{U.~G.}
  \bibnamefont{Meissner}}, \bibinfo{journal}{Nucl. Phys.}
  \textbf{\bibinfo{volume}{A635}}, \bibinfo{pages}{121} (\bibinfo{year}{1998}),
  \eprint{hep-ph/9801297}.

\bibitem[{\citenamefont{Hagler}(2009)}]{Hagler:2009ni}
\bibinfo{author}{\bibfnamefont{P.}~\bibnamefont{Hagler}}
  (\bibinfo{year}{2009}), \eprint{0912.5483}.

\bibitem[{\citenamefont{Bernard et~al.}(1991)\citenamefont{Bernard, Kaiser, and
  Meissner}}]{Bernard:1991rq}
\bibinfo{author}{\bibfnamefont{V.}~\bibnamefont{Bernard}},
  \bibinfo{author}{\bibfnamefont{N.}~\bibnamefont{Kaiser}}, \bibnamefont{and}
  \bibinfo{author}{\bibfnamefont{U.~G.} \bibnamefont{Meissner}},
  \bibinfo{journal}{Phys. Rev. Lett.} \textbf{\bibinfo{volume}{67}},
  \bibinfo{pages}{1515} (\bibinfo{year}{1991}).

\bibitem[{\citenamefont{Bernard et~al.}(1993)\citenamefont{Bernard, Kaiser,
  Schmidt, and Meissner}}]{Bernard:1993bg}
\bibinfo{author}{\bibfnamefont{V.}~\bibnamefont{Bernard}},
  \bibinfo{author}{\bibfnamefont{N.}~\bibnamefont{Kaiser}},
  \bibinfo{author}{\bibfnamefont{A.}~\bibnamefont{Schmidt}}, \bibnamefont{and}
  \bibinfo{author}{\bibfnamefont{U.~G.} \bibnamefont{Meissner}},
  \bibinfo{journal}{Phys. Lett.} \textbf{\bibinfo{volume}{B319}},
  \bibinfo{pages}{269} (\bibinfo{year}{1993}), \eprint{hep-ph/9309211}.

\bibitem[{\citenamefont{Hemmert et~al.}(1997)\citenamefont{Hemmert, Holstein,
  and Kambor}}]{Hemmert:1996rw}
\bibinfo{author}{\bibfnamefont{T.~R.} \bibnamefont{Hemmert}},
  \bibinfo{author}{\bibfnamefont{B.~R.} \bibnamefont{Holstein}},
  \bibnamefont{and} \bibinfo{author}{\bibfnamefont{J.}~\bibnamefont{Kambor}},
  \bibinfo{journal}{Phys. Rev.} \textbf{\bibinfo{volume}{D55}},
  \bibinfo{pages}{5598} (\bibinfo{year}{1997}), \eprint{hep-ph/9612374}.

\bibitem[{\citenamefont{Beane et~al.}(2005)\citenamefont{Beane, Malheiro,
  McGovern, Phillips, and van Kolck}}]{Beane:2004ra}
\bibinfo{author}{\bibfnamefont{S.~R.} \bibnamefont{Beane}},
  \bibinfo{author}{\bibfnamefont{M.}~\bibnamefont{Malheiro}},
  \bibinfo{author}{\bibfnamefont{J.~A.} \bibnamefont{McGovern}},
  \bibinfo{author}{\bibfnamefont{D.~R.} \bibnamefont{Phillips}},
  \bibnamefont{and} \bibinfo{author}{\bibfnamefont{U.}~\bibnamefont{van
  Kolck}}, \bibinfo{journal}{Nucl. Phys.} \textbf{\bibinfo{volume}{A747}},
  \bibinfo{pages}{311} (\bibinfo{year}{2005}), \eprint{nucl-th/0403088}.

\bibitem[{\citenamefont{Edwards and Joo}(2005)}]{Edwards:2004sx}
\bibinfo{author}{\bibfnamefont{R.~G.} \bibnamefont{Edwards}} \bibnamefont{and}
  \bibinfo{author}{\bibfnamefont{B.}~\bibnamefont{Joo}}
  (\bibinfo{collaboration}{SciDAC}), \bibinfo{journal}{Nucl. Phys. Proc.
  Suppl.} \textbf{\bibinfo{volume}{140}}, \bibinfo{pages}{832}
  (\bibinfo{year}{2005}), \eprint{hep-lat/0409003}.

\bibitem[{\citenamefont{Einstein and Laub}(1908)}]{Einstein:1906aa}
\bibinfo{author}{\bibfnamefont{A.}~\bibnamefont{Einstein}} \bibnamefont{and}
  \bibinfo{author}{\bibfnamefont{J.}~\bibnamefont{Laub}},
  \bibinfo{journal}{Annalen d.~Phys.} \textbf{\bibinfo{volume}{331}},
  \bibinfo{pages}{532} (\bibinfo{year}{1908}).

\end{thebibliography}

\end{document}